\documentclass[aps,pra,reprint,amsmath,superscriptaddress,amssymb,showpacs,nofootinbib, onecolumn]{revtex4-2}
\usepackage{bbm, bm, mathtools, braket, graphicx, hyperref, graphicx, mathptmx, xcolor,amsmath,amssymb}
\usepackage[utf8]{inputenc}
\usepackage[T1]{fontenc}
\usepackage{tikz}
\usetikzlibrary{calc}
\usetikzlibrary{decorations}
\usetikzlibrary{plotmarks}

\newcommand{\op}[1]{\hat{#1}}
\newcommand{\Id}{\op{\mathbbm{1}}}
\newcommand{\Tr}{\mathrm{Tr}}
\newcommand{\LParen}{ \bm{(} }
\newcommand{\RParen}{ \bm{)} }

\begin{document}

\title{Integrability of Goldilocks quantum cellular automata}

\author{Logan E. Hillberry }
\affiliation{Department of Physics, The University of Texas at Austin, Austin, TX 78712, USA}
\email[]{lhillber@utexas.edu }
\affiliation{JILA, NIST and University of Colorado, Department of Physics, University of Colorado, Boulder, Colorado 80309, USA}
\author{Lorenzo Piroli }
\affiliation{Dipartimento di Fisica e Astronomia, Universit\`a di Bologna and INFN, Sezione di Bologna, via Irnerio 46, I-40126 Bologna, Italy}
\author{Eric Vernier}
\affiliation{Laboratoire de Probabilités, Statistique et Modélisation CNRS, Université Paris Cité, Sorbonne Université Paris, France}
\author{Nicole Yunger Halpern }
\affiliation{Joint Center for Quantum Information and Computer Science, NIST and University of Maryland, College Park, MD 20742, USA}
\affiliation{Institute for Physical Science and Technology, University of Maryland, College Park, MD 20742, USA}
\author{Toma\v{z} Prosen }
\affiliation{Faculty of Mathematics and Physics, University of Ljubljana, Jadranska 19, SI-1000 Ljubljana, Slovenia }
\affiliation{Institute of Mathematics, Physics, and Mechanics, Jadranska 19, SI-1000 Ljubljana, Slovenia}
\author{Lincoln D. Carr }
\affiliation{Quantum Engineering Program, Colorado School of Mines, Golden, CO 80401, USA }
\affiliation{Department of Physics, Colorado School of Mines, Golden, CO 80401, USA }
\affiliation{Department of Applied Mathematics and Statistics, Colorado School of Mines, Golden, CO 80401, USA 
}
\date{\today}

\begin{abstract}
Goldilocks quantum cellular automata (QCA) have been simulated on quantum hardware and produce emergent small-world correlation networks. In Goldilocks QCA, a single-qubit unitary is applied to each qubit in a one-dimensional chain subject to a balance constraint: a qubit is updated if its neighbors are in different computational-basis states. We prove that a subclass of Goldilocks QCA, including the QCA implemented experimentally, map to free fermions and therefore can be simulated classically. We support this claim with two proofs, one involving a Jordan--Wigner transformation and one mapping the integrable six-vertex model to QCA. We compute local conserved quantities of these QCA and predict experimentally measurable expectation values. These calculations can be applied to test large digital quantum computers. In contrast, typical Goldilocks QCA have equilibration properties and quasienergy-level statistics that suggest nonintegrability. Still, each of the latter QCA conserves one quantity useful for error mitigation. Our work yields a parametric quantum circuit with tunable integrability properties useful for testing quantum hardware.
\end{abstract}

\maketitle

\section{Introduction}

Classical cellular automata (CA) are dynamical models that evolve bit strings (sequences of $1$s and $0$s) under simple local rules~\cite{delorme1999introduction}. At each time step, every bit is updated in accordance with a rule and its neighbors' states. Despite their simplicity, CA engender rich dynamics: order, randomness, fractals, and conservation laws. Also, CA can implement universal classical computation~\cite{wolfram1983statistical,cook2004universality}.  Early CA work stressed conservation laws as ingredients for modeling physics~\cite{margolus1987physics, adler1995simulating}.

Classical CA generalize to quantum cellular automata (QCA)~\cite{arrighi2019overview,farrelly2020review}. QCA evolve under discrete dynamics implementable on quantum computers with local gates and, in some cases, ancillas. We focus on a class of QCA that we call \emph{digital QCA}: constrained models implementable with quantum circuits of geometrically local gates. Examples include \emph{Goldilocks QCA}, which generate small-world networks of bipartite mutual information~\cite{hillberry2021entangled, jones2022small}, and the \emph{Floquet PXP model}~\cite{wilkinson2020exact,prosen2021many}, which exhibits many-body scars~\cite{turner2018weak,wang2024quantum}. As local, unitary, discrete models, digital QCA parallel reversible classical CA~\cite{hillberry2021entangled,casagrande2023complexity,lesanovsky2019non,wintermantel2020unitary}.

Quantum hardware recently simulated digital QCA~\cite{jones2022small}. The noisy data were postselected to obey a conservation law. Consequently, complex networks of two-qubit correlators survived $>1,000$ two-qubit gates. An important open question is whether classical computers can efficiently simulate any digital QCA. (Throughout the rest of this paper, ``classically simulable'' implicitly means ``efficiently classically simulable.'') Identifying classically simulable quantum models is important: on the one hand, it restricts the set of models that allow for practical quantum advantages~\cite{daley2022practical}. On the other hand, classically simulable models enable tests of quantum computers. Quantum hardware likely will soon (i) contain more qubits than one can simulate by representing the state vector fully and (ii) generate more entanglement than tensor-network methods can accommodate. Classically simulable models will enable comparisons between such hardware and theoretical predictions. Meanwhile, experimental realization of nonintegrable dynamics could demonstrate quantum advantage.

Many physical systems conserve quantities such as energy, momentum, and particle number. Informally, a classical system is \emph{integrable} if it conserves enough quantities that it evolves predictably (one can solve the dynamics exactly). Two-body gravitational orbits and harmonic oscillators exemplify classical integrabilty.  Even discrete integrable systems obey conservation laws: in classical and quantum statistical physics, exactly solvable discrete systems include the two-dimensional Ising model~\cite{onsager1944crystal} and the six-vertex model of ice~\cite{pauling1935structure,lieb1967residual}.

Although integrable quantum systems conserve nontrivial properties, defining integrability remains a research challenge~\cite{caux2011remarks}. For example, some conserved properties are trivial: every Hamiltonian conserves its eigenprojectors, which therefore cannot qualify the system for integrability. Requiring locality of the conserved quantities excludes the eigenprojectors, but quasilocal quantities introduce further complications~\cite{prosen2011open}. Some one-dimensional quantum models~\cite{jordan1928uber,bethe1931theorie,yang1967some,baxter1972one} are integrable in that (i) the dynamics conserve extensively many local quantities and (ii) many-body interactions decompose into two-body interactions whose chronological ordering does not affect the physics. \emph{Yang–Baxter integrability} codifies these two points, exemplifying integrability~\cite{caux2011remarks}. The six-vertex and two-dimensional Ising models can encode (1+1)-dimensional quantum dynamics that are integrable in the Yang–Baxter sense. Yang–Baxter integrability thereby interrelates classical and quantum-lattice integrability \cite{baxter1972one,baxter2008exactly,retore2022introduction,bazhanov2023ising}. More recently, researchers have extended the notion of integrability to quantum circuits~\cite{vanicat2018integrable,ljubotina2019ballistic,bertini2019exact,medenjak2020rigorous, claeys2022correlation,vernier2023integrable,Miao2023integrablequantum,vernier2024strong,fukai2025quantum,szasz2025construction}. 

Classical computers can simulate some integrable quantum systems. The simplest quantum-integrability setting involves noninteracting, or \emph{free}, dynamics. Classical computers can simulate free fermions prepared in Gaussian states~\cite{terhal2002classical,bravyi2004lagrangian,surace2022fermionic}. The Jordan-Wigner (JW) transformation can map qubits to fermions~\cite{jordan1928uber,lieb1961two,cabrera1987role,li2022higher}. However, it is nontrivial to identify whether a unitary-gate sequence maps to a noninteracting one.\footnote{An elegant graph-theoretic solution for identifying Hamiltonians that generate noninteracting dynamics appeared recently~\cite{chapman2020characterization}. However, not all Hamiltonians with free-fermionic spectra can be cast, via JW mappings, as sums of terms bilinear in fermionic operators~\cite{fendley2019free,alcaraz2020free,elman2021free,chapman2023unified,fendley2023free,vernier2025hilbert}.} Classical computers cannot necessarily simulate (i) integrable interacting systems or (ii) free fermions prepared in non-Gaussian states.

We identify Goldilocks QCA that map to free fermions, presenting two proofs. One relies on a JW transformation; and one, on a map from the integrable six-vertex model. This work's novelty stems partially from the lack of any systematic procedure for identifying noninteracting-fermionic QCA. Furthermore, we compute conserved quantities (\emph{charges}) and use them to predict time-averaged expectation values. Unlike their free-fermionic counterparts, generic Goldilocks QCA appear nonintegrable according to three lines of evidence: each conserves only one local charge, local observables' expectation values equilibrate to values predictable from that charge alone, and each QCA displays quasienergy-level statistics consistent with quantum chaos. In summary, we present a parametric model that has tunable integrability properties and tunable conservation laws. Our work elevates digital QCA as a practical tool for benchmarking large quantum computers. For instance, one could implement our integrable QCA and measure charges' expectation values at various times. Any experimentally observed violation of conservation would imply an experimental error.

\begin{figure}
\includegraphics[]{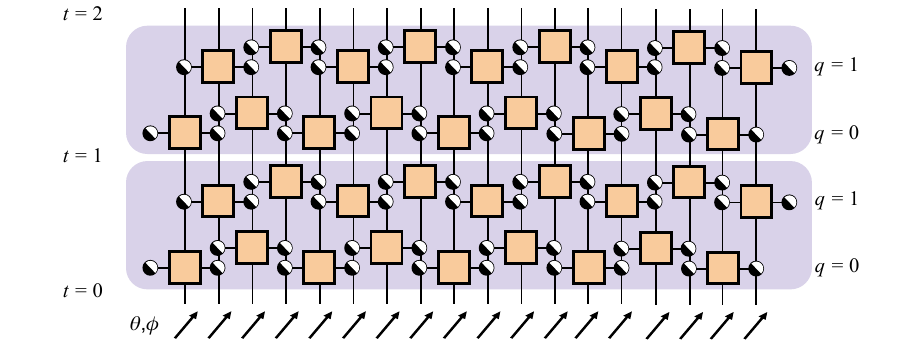}
\caption{ \emph{Goldilocks-QCA circuit.} 
    Each half-black-half-white circle represents one control of a generalized XOR constraint. The orange box represents the single-qubit unitary $\op{V}$. 
    The initial state is a tilted ferromagnet with a polar angle $\theta$ and an azimuthal angle $\phi$. Each dangling control gate wraps around to the other side of the system, obeying periodic boundary conditions. The purple box marks the time-step gate $\op{U}$, which consists of two layers, $\op{G}(\op{V}, q)$, wherein $q = 0, 1$.
}
\label{fig:schematic}
\end{figure}

The rest of this work is organized as follows. Section~\ref{sec:the_model} introduces digital QCA. Section~\ref{sec:integrable_QCA} contains our main results. There, we identify the QCA for which we can prove a free-fermionic mapping (Sections~\ref{sec:mapping_JW} and ~\ref{sec:mapping_six_v_proof}). Also, we exhibit a family of conservation laws (Sec.~\ref{sec:identify_charges}). Section~\ref{sec:simulation_dynamics} presents our simulations of QCA dyanmaics and numerical results of level statistics. Section~\ref{sec:outlook} relates our conclusions.

\section{The model}
\label{sec:the_model}

We study a one-dimensional chain of qubits indexed by $j=0,1,\ldots, L-1$. Qubit $j$ corresponds to a Hilbert space $\mathcal{H}_j = \mathbb{C}^2$; and the chain, to $\mathcal{H}=\otimes_j \mathcal{H}_j$. We denote by $\{ \ket{0}, \ket{1} \}$ the single-qubit computational basis and by $\op{\sigma}^\gamma_j$ the $\gamma\in\{x,y,z\}$ Pauli operator of qubit $j$. We denote the identity by $\Id_j$. For convenience, we choose periodic boundary conditions, $j+L\equiv j$, and an even number $L$ of qubits. Our results do not depend on these choices. 

$\ket{\Psi_t}$ denotes the system's state at the discrete time $t$. $\ket{\Psi_t}$ updates via spacetime-periodic applications of a local controlled-unitary \emph{neighborhood gate}. The gate is applied in a brickwork pattern to all even-index qubits, then all odd-index ones (Fig.~\ref{fig:schematic}). During each time step, a single-qubit gate $\op{V} \in {\rm U}(2)$ may be applied to each qubit, depending on the states of that qubit's neighbors.
%
%\footnote{Since other qubits' states control the application of $\op{V}$, the operator's phase affects the global evolution. Therefore, $\det (\op{V})$ need not equal one; generally, $\op{V} \in {\rm U(2)}$. We sidestep this subtlety, restricting to $\op{V} \in {\rm SU(2)}$.} 
%
The neighborhood gates $\op{u}_j(\op{V})$ have the form
\begin{equation} \label{eq:uj}
	\op{u}_j(\op{V}) \coloneqq \sum_{m,n=0}^1 \ket{m} \! \bra{m}_{j-1} \otimes (\op{V}_j)^{f(m,n)}\otimes \ket{n} \! \bra{n}_{j+1}\, .
\end{equation}
The left- and right-neighboring qubits form the controls. $f(m,n) \in \{0,1\}$ encodes the local update rule: if $f(m,n)=1$ while qubit $j$'s neighbors are in $\ket{m,n}_{j-1,j+1}$, then $\op{V}$ updates qubit $j$'s state; otherwise, not. The global even-site/odd-site half-time-step gates are 
\begin{equation}\label{eq:two-step_evolution}
	\op{G}{(\op{V},q)} \coloneqq \prod_{k=0}^{L/2-1} \op{u}_{2k+q}(\op{V})\, , \quad q=0,1\, .
\end{equation}
A unitary $\op{U}$ defined on the Hilbert space $\mathcal{H}$ describes the dynamics: $\ket{\Psi_{t+1}} = \op{U}(\op{V}) \ket{\Psi_t}$, wherein $\op{U}(\op{V}) \coloneqq \op{G}{(\op{V},1)} \, \op{G}{(\op{V},0)}$. 

Digital QCA differ in their rule $f$ and local unitary $\op{V}$. For example, the Floquet PXP model flips qubit $j$ if the neighbors are in $\ket{00}_{j-1, j+1}$~\cite{wilkinson2020exact,prosen2021many}. The corresponding $\op{V}=\op{\sigma}^x$, and $f_{\rm PXP}(m,n)=1-m \oplus n - mn$. The $\oplus$ denotes addition modulo $2$, the exclusive-or (XOR) operation. The Floquet Fredrickson-Andersen model flips a qubit if its neighbors are in $\ket{01}_{j-1, j+1}$, $\ket{10}_{j-1, j+1}$, or $\ket{11}_{j-1, j+1}$~\cite{gopalakrishnan2018facilitated,gopalakrishnan2018operator}. This model's $\op{V}=\op{\sigma}^x$, and $f_{\rm FA}(m,n)= m \oplus n + mn$. 
For Goldilocks QCA,
\begin{equation}\label{eq:goldilocks_rule}
    f_{\rm G}(m,n)=m\oplus n\, ;
\end{equation}
the neighbor configurations $\ket{01}_{j-1, j+1}$ and $\ket{10}_{j-1, j+1}$ enable nontrivial evolution~\cite{hillberry2021entangled}. The name \emph{Goldilocks} QCA comes from the rule's updating qubit $j$ under more neighborhood conditions than the PXP model but fewer conditions than the Fredrickson-Andersen model---under a ``just right'' balance of activity and inactivity~\cite{hillberry2021entangled}. When $\op{V}=\op{\sigma}^x$, Goldilocks QCA have been called the XOR-Fredrickson-Andersen model~\cite{causer2020dynamics} and the XOR QCA \cite{arrighi2008one}. Our digital QCA generalize these previously studied models and certain reversible CA. $\op{V}$ introduces extra degrees of freedom, as the most general single-site unitary is, to within a global phase,
\begin{equation} \label{eq:possibly_non_integrable}
    \op{V}(a, b) \coloneqq \frac{1}{\sqrt{|a|^2 +|b|^2}} \left(
    \begin{array}{cc}
        a & b \\
        -b^* & a^*
    \end{array}
\right)\, ,
\quad a, b \in \mathbb{C} \, .
\end{equation}
As we will show in the next section, different $\op{V}$s lead to integrable and seemingly nonintegrable dynamics. Henceforth, we denote by $\op{U}(\op{V})$ the \emph{Goldilocks}-QCA time-step gate constructed with the gate $\op{V}$.

\section{Two proofs that certain Goldilocks QCA are integrable}
\label{sec:integrable_QCA}

This section contains our main result: the identification of the Goldilocks QCA that map to free-fermionic (and hence exactly solvable) dynamics. Define the single-site update unitary
\begin{equation} \label{eq:V_free}
    \op{V}_{\rm free}(\alpha, \beta, \pm) \coloneqq 
    \left(
\begin{array}{cc}
\cos\alpha &  \mp e^{-i\beta}\sin\alpha \\ 
 e^{i\beta} \sin\alpha &  \pm \cos\alpha
\end{array}
\right) 
    \,,
\enspace \alpha, \beta \in \mathbb{R}.
\end{equation}
One can interpret the parameters in terms of single-qubit rotations. The operator for rotation by $\theta$ about $\gamma$ is $\op{\mathcal{R}}_{\gamma}(\theta):=e^{-i (\theta/2) \op{\sigma}^{\gamma}}$. Any single-qubit unitary can be decomposed into a global phase factor times an Euler-angle-rotation sequence $\op{R}_{z}(\theta_3) \op{R}_{y}(\theta_2) \op{R}_{z}(\theta_1)$. The single-site update unitary $\op{V}_{\rm free}(\alpha, \beta,\pm)$ decomposes into different Euler-angle sequences for the different sign choices $(\pm)$: $\op{V}_{\rm free}(\alpha, \beta, +)=\op{\mathcal{R}}_{z}(\beta) \op{\mathcal{R}}_{y}(2 \alpha) \op{\mathcal{R}}_z(-\beta)$, and $\op{V}_{\rm free}(\alpha, \beta, -)=i\op{\mathcal{R}}_{z}(\beta) \op{\mathcal{R}}_{y}(2 \alpha) \op{\mathcal{R}}_z(-\beta+\pi)$. We show that $\op{U}(\op{V}_{\rm free})$ generates free-fermionic dynamics. In the next two subsections, we prove this fact in two ways.

\subsection{Mapping of certain Goldilocks QCA to free fermions via JW transformation}
\label{sec:mapping_JW}

We now prove via our first strategy that the unitary~\eqref{eq:V_free} gives rise to free-fermionic dynamics. We will define Hamiltonians that are bilinear in the fermion creation ($\op{a}_j^{\dagger}$) and annihilation ($\op{a}_j$) operators. These Hamiltonians generate the global time-step unitaries $\op{U}\bm{(}\op{V}_{\rm free}(\alpha, \beta, \pm)\bm{)}$. We first prove the result for restricted arguments $\op{V}_{\rm free}(\alpha,0,-)$, then generalize. 

Our proof relies on the JW transformation\footnote{The most common JW transformation reads $\op{a}_j=\prod_{k<j} \op{\sigma}^z_k (\op{\sigma}^x_j+i\op{\sigma}^y_j)/2$. Our choice renders the gates $\op{\sigma^y}$ and $\op{\sigma}^z_j  \op{\sigma}^z_{j+1}$ free-fermionic, as necessary for our proof. The usual results hold under the cyclic replacement ``usual JW'' $\mapsto$ ``our JW'': $z \mapsto y$, $x \mapsto z$, and $y \mapsto x$. Our JW transformation privileges the $y$-direction, rather than the conventional $z$-direction, due to the $e^{\pm i (\alpha/2) \op{\sigma}^y_j}$ in $\op{u}_j$.}
\begin{align}\label{eq:fermionic}
    \op{a}_j \coloneqq \prod_{k=0}^{j-1} \op{\sigma}^y_k \, \frac{\op{\sigma}^z_j+i\op{\sigma}^x_j}{2}\, ,
\end{align}
defining fermionic operators satisfying canonical commutation relations.
The transformation~\eqref{eq:fermionic} does not map the neighborhood gates~\eqref{eq:uj} to Gaussian fermionic operators, so the fact that the QCA map to free fermions is nontrivial. Contrariwise, matchgate circuits are free fermionic because they are compositions of local free-fermionic gates~\cite{terhal2002classical,vanDenNest2011simulating,brod2016efficient}. Despite this lack of precedence, we will show that~\eqref{eq:fermionic} maps the global unitaries $\op{U}\bm{(}\op{V}_{\rm free}(\alpha, \beta, \pm)\bm{)}$ to Gaussian operators. 

Consider the neighborhood gate $\op{u}_j$ [Eq.~\eqref{eq:uj}] with the restricted argument $\op{V}_{\rm free}(\alpha, 0, -)$.
This gate decomposes into single-qubit rotations about the $y$-axis and controlled-$z$ gates, denoted $[CZ]_{i,j} = \ket{0}\bra{0}_i \otimes \Id_j + \ket{1}\bra{1}_i \otimes \op{\sigma}^z_j$ and for which $[CZ]_{i,j}=[CZ]_{j,i}$:
\begin{equation}\label{eq:split_v}
	\op{u}_j \bm{(} \op{V}_{\rm free}(\alpha, 0, -) \bm{)}
    =e^{-i (\alpha/2) \,
    \op{\sigma}^y_j} \,
    [C Z]_{j-1, j} \,
    [C Z]_{j, j+1} \,
    e^{i (\alpha/2) \op{\sigma}^y_j} \, .
\end{equation}
One can check this decomposition by direct computation.
In terms of these neighborhood gates, the even ($q=0$) and odd ($q=1$) half-time-step operators have the forms 
\begin{align} 
\op{G}\bm{(}\op{V}_{\rm free}(\alpha,0,-),\, q\bm{)}
& = \prod_{j=0}^{L / 2-1} \op{u}_{2j+q} \bm{(} \op{V}_{\rm free}(\alpha, 0, -) \bm{)} \nonumber\\
& = \left ( \prod_{j=0}^{L/2 - 1} e^{-i (\alpha/2) \op{\sigma}_{2j+q}^y} \right )
   \left ( \prod_{j=0}^{L/2 - 1}[CZ]_{j-1,j}[CZ]_{j,j+1} \right )
   \left( \prod_{j=0}^{L/2 - 1} e^{i (\alpha/2) \op{\sigma}_{2j+q}^y} \right ). \label{eq:split_uj}
\end{align}
The controlled-$z$ gate decomposes in terms of Pauli operators as
\begin{equation}\label{eq:cz}
   [C Z]_{j, j+1}
   =e^{-i \pi/4} \,
   \exp \left(-i \frac{\pi}{4} \, \op{\sigma}^z_j \,
               \op{\sigma}^z_{j+1}\right) \,
   \exp \left( +i \, \frac{\pi}{4}
   \left[ \op{\sigma}^z_j+\op{\sigma}^z_{j+1}\right] \right).
\end{equation}
Therefore, the second factor in Eq.~\eqref{eq:split_uj} decomposes similarly as
\begin{align} 
   \prod_{j=0}^{L/2 - 1}[CZ]_{j-1,j}[CZ]_{j,j+1} &= \prod_{j=0}^{L - 1}[CZ]_{j,j+1} 
   % % %
   = \prod_{j=0}^{L-1}\exp \left(-i \frac{\pi}{4} \op{\sigma}^z_j \op{\sigma}^z_{j+1}\right) 
   \exp \left(i \frac{\pi}{4}\left[\op{\sigma}^z_j+\op{\sigma}^z_{j+1}\right]\right) \nonumber \\
   % % %
   & =\exp\left(-i\frac{\pi}{4}\sum_{j=0}^{L-1}\op{\sigma}^z_j\op{\sigma}^z_{j+1}\right) 
   \exp\left(i\frac{\pi}{2}\sum_{j=0}^{L-1}\op{\sigma}^z_j\right)
   \nonumber\\
   % % % 
   & = i^L\exp\left(-i\frac{\pi}{4}\sum_{j=0}^{L-1}\op{\sigma}^z_j\op{\sigma}^z_{j+1}\right) \prod_{k=0}^{L-1} \op{\sigma}^z_{k}\,. \label{eq:global_cz}
\end{align}
The final equality follows from $e^{i(\pi/2) \op{\sigma}^z}=i\op{\sigma}^z$. Similarly,
$\op{\sigma}^z_k\op{\sigma}^z_{k+1}
    =-i \exp\left(i\frac{\pi}{2} \op{\sigma}^z_k \op{\sigma}^z_{k+1}\right)$, 
so the final factor in~\eqref{eq:global_cz} is
\begin{align}
    \prod_{k=0}^{L-1} \op{\sigma}^z_{k} 
     = \prod_{k=0}^{L/2-1} \op{\sigma}^z_{2k} \op{\sigma}^z_{2k+1}
     = (-i)^{L/2} \exp\left(i\frac{\pi}{2}\sum_{j=0}^{L/2-1}\op{\sigma}^z_{2k}\op{\sigma}^z_{2k+1}\right). \label{eq:global_z}
\end{align}
We substitute from Eq.~\eqref{eq:global_z} into Eq.~\eqref{eq:global_cz}, then substitute the result into Eq.~\eqref{eq:split_uj}. Neglecting the global phase, we obtain
\begin{align}
   \label{eq:G_help1}
   \op{G}\bm{(}\op{V}_{\rm free}(\alpha,0,-),\, q\bm{)}
   =\exp \left(\frac{-i \alpha}{2} \sum_{j=0}^{L / 2-1} \op{\sigma}^y_{2 j+q}\right) 
   \exp \left( \frac{-i \pi}{4} 
     \left\{ \op{\sigma}^z_{L-1} \op{\sigma}^z_0
     +\sum_{j=0}^{L-2}[-1]^{j+1} \op{\sigma}^z_j \op{\sigma}^z_{j+1}\right\} \right) 
   \exp \left( \frac{i \alpha}{2} \sum_{k=0}^{L / 2-1} \op{\sigma}^y_{2 j+q}\right) .
\end{align}
Inspired by the exponentials' arguments, we define the Hamiltonians
\begin{align}
   \label{H12_app}
   & \op{H}_1(\alpha,\,q) \coloneqq -\frac{\alpha}{2} \sum_{j=0}^{L / 2-1} \op{\sigma}^y_{2 j+q}
   \quad \text{and} \nonumber\\
   % % %
   & \op{H}_2 \coloneqq \frac{\pi}{4}\left[ \op{\sigma}^z_{L-1} \op{\sigma}^z_0+\sum_{j=0}^{L-2}(-1)^{j+1} \op{\sigma}^z_j \op{\sigma}^z_{j+1}\right] \, .
\end{align}
Substituting into Eq.~\eqref{eq:G_help1} yields the following formula for the half-time-step operator:
\begin{equation}
    \op{G}\bm{(}\op{V}_{\rm free}(\alpha,0,-),\,q\bm{)}=\exp \bm{(}+i H_1(\alpha,q) \bm{)} \exp (-i H_2) \exp \bm{(} -i H_1(\alpha, q) \bm{)}\,.
\end{equation}

Let us replace the Pauli operators, using the JW transformation~\eqref{eq:fermionic}, with fermion creation and annihilation operators. From the latter, we can build bilinear fermionic operators such as $\op{\sigma}^y_j=\op{a}_{j} \op{a}_{j}^{\dagger}-\op{a}_{j}^{\dagger} \op{a}_{j}$ and $\op{\sigma}^z_{j}\op{\sigma}^z_{j+1} = (\op{a}_j^{\dagger}-\op{a}_j)(\op{a}_{j+1}+\op{a}_{j+1}^{\dagger})$.
Using the $\op{\sigma}^y_j$ formula, we construct the fermion-number-parity operator,
\begin{equation}
    \op{\mathcal{P}} = \prod_{j=0}^{L-1}\op{\sigma}^y_{j} 
    = \prod_{j=0}^{L-1} \left( \op{a}_{j} \op{a}_{j}^{\dagger}-\op{a}_{j}^{\dagger} \op{a}_{j} \right)\,.
\end{equation}
The number-parity operator features in one of the Hamiltonians we defined: applying the JW transformation~\eqref{eq:fermionic} to Eqs.~\eqref{H12_app} yields
\begin{align}   \label{eq:H1}
   & \op{H}_1(\alpha, q)=-\frac{\alpha}{2} \sum_{j=0}^{L / 2-1}\left(\op{a}_{2 j+q} \op{a}_{2 j+q}^{\dagger}-\op{a}_{2 j+q}^{\dagger} \op{a}_{2 j+q}\right)\,
   \quad \text{and} \\
   % % %
   \label{eq:H2_as}
   & \op{H}_2 =\frac{\pi}{4}\left[ \left(\op{a}_0+\op{a}_0^{\dagger}\right)\left(\op{a}_{L-1}-\op{a}_{L-1}^{\dagger}\right) \op{\mathcal{P}}+\sum_{j=0}^{L-2}(-1)^{j+1}\left(\op{a}_j^{\dagger}-\op{a}_j\right)\left(\op{a}_{j+1}+\op{a}_{j+1}^{\dagger}\right)\right].
\end{align}

$\op{H}_1$ is overtly quadratic in the fermionic operators, whereas $\op{H}_2$ is not . Yet we can prove that $\op{H}_2$ is quadratic by invoking the number-parity operator. This operator commutes with both Hamiltonians:
$[\op{H}_2, \op{\mathcal{P}}]=[\op{H}_1(\alpha,q), \op{\mathcal{P}}] = 0$.
Each $\op{H}$ can be therefore represented, relative to the $\op{\mathcal{P}}$ eigenbasis, by a block-diagonal matrix. 
Each matrix consists of two blocks, as $\op{\mathcal{P}}$ has two eigenvalues, $\pm 1$. Denote the $j^{\rm th}$ $+1$ eigenstate by $\ket{\varphi_+^{(j)} }$ and the $k^{\rm th}$  $-1$ eigenstate by $\ket{\varphi_-^{(k)} }$. In terms of these eigenstates, an arbitrary pure state decomposes as
$\ket{\Psi} = \sum_j c_+^{(j)} \ket{\varphi_+^{(j)} } 
+ \sum_k c_-^{(k)} \ket{\varphi_-^{(k)} }$,
wherein the coefficients $c_+^{(j)}, c_-^{(k)} \in \mathbb{C}$ are normalized. Consider operating with $\op{H}_2$ on $\ket{\Psi}$. Whenever the operator acts on a $\ket{\varphi_+^{(j)} }$ term, the $\op{\mathcal{P}}$ in Eq.~\eqref{eq:H2_as} acts as $+1$; and, when $\op{H}_2$ acts on a $\ket{\varphi_-^{(k)} }$ term, the $\op{\mathcal{P}}$ acts as a $-1$. Hence, $\op{H}_2$ always acts as though a $\pm 1$ replaced the $\op{\mathcal{P}}$ in Eq.~\eqref{eq:H2_as}:
\begin{align}
     \op{H}_2 &\coloneqq \frac{\pi}{4}\left[\pm \left(\op{a}_0+\op{a}_0^{\dagger}\right)
    \left(\op{a}_{L-1}-\op{a}_{L-1}^{\dagger}\right)  + \sum_{j=0}^{L-2}(-1)^{j+1}
    \left(\op{a}_j^{\dagger}
    -\op{a}_j\right)
    \left(\op{a}_{j+1}+\op{a}_{j+1}^{\dagger}\right)\right]\,. \label{eq:H2}
\end{align}
This formula is overtly quadratic in the fermionic operators.

Therefore, both half-time-step operators $\op{G}(\op{V}_{\rm free},\,q)$, $q\in\{0,1\}$ are quadratic and so free-fermionic. 
Consequently, the global-evolution operator $\op{U}(\op{V}_{\rm free})$, which is a product of the two half-time-step operators, has the form
\begin{equation}\label{eq:app_final_factorization}
   \op{U}\bm{(} \op{V}_{\rm free}(\alpha,0,-) \bm{)}
   =e^{i\op{H}_1(\alpha,\, 1)} \, e^{-i\op{H}_2} \,
   e^{-i\op{H}_1(\alpha,\, 1)} \, e^{i\op{H}_1(\alpha,\,0)} \,
   e^{-i\op{H}_2} \, e^{-i\op{H}_1(\alpha,\,0)}\,.
\end{equation}
Since $\op{H}_{1,2}$ are free-fermionic, the Goldilocks-QCA time-step operator effects free-fermionic dynamics. Hence the Goldilocks QCA~\eqref{eq:app_final_factorization} is integrable. 

One could further simplify Eq.~\eqref{eq:app_final_factorization} to compute the Floquet Hamiltonian $\op{K}$ that satisfies $\op{U}=e^{-i\op{K}}$. One would Fourier-transform Eq.~\eqref{eq:app_final_factorization}, then apply the Baker–Campbell–Hausdorff formula. Finally, one would invoke the fact that any two quadratic operators’ commutator is a quadratic operator. See Ref.~\cite{vernier2023integrable} for a similar computation. However, we do not use the form of $\hat{K}$, so we omit this computation. 

We now relax the restrictions imposed on the arguments of $\op{V}_{\rm free}$ [Eq.~\eqref{eq:V_free}], the $\beta$ and then the minus sign. The already-analyzed operator $\op{V}(\alpha,0,-)$ transforms into the less restricted operator
$\op{V}_{\rm free}(\alpha,\beta, -)$ 
under the single-site unitary
$e^{-i(\beta/2)\op{\sigma}^z}$:
$\op{V}_{\rm free}(\alpha,\beta, -)
=e^{-i(\beta/2)\op{\sigma}^{z}}
\op{V}(\alpha,0,-)
e^{i(\beta/2)\op{\sigma}^{z}}$. 
The change of basis preserves the QCAs' free-fermionic structure, merely rotating the Pauli operators in  the JW transformation~\eqref{eq:fermionic}:
\begin{equation}\label{eq:b_replacements}
    \op{\sigma}^x \mapsto \cos \beta\, \op{\sigma}^x + \sin \beta\, \op{\sigma}^y\, , 
    \quad \text{and} \quad
    \op{\sigma}^y \mapsto \sin\, \beta \op{\sigma}^x + \cos\, \beta \op{\sigma}^y\, .
\end{equation}

Finally, we prove that $\op{V}_{\rm free}(\alpha, \beta, +)$ is free-fermionic. Direct calculation reveals 
\begin{equation}
u_j\bm{(}\op{V}_{\rm free}(\alpha, \beta, +)\bm{)} = [CZ]_{j-1,j}[CZ]_{j,j+1}u_j\bm{(}\op{V}_{\rm free}(-\alpha, \beta, -)\bm{)}\,.    
\end{equation}
Therefore, one can simulate $\op{V}_{\rm free}(\alpha, \beta, +)$ using $\op{V}_{\rm free}(-\alpha, \beta, -)$: one flips the sign of $\alpha$ and appends a layer of controlled-$z$ gates: 
\begin{equation} \label{eq:Vfree_plus}
    \op{G}\bm{(}\op{V}_{\rm free}(\alpha, \beta, +),\, q\bm{)} = \prod_{j=0}^{L-1} [CZ]_{j,j+1}  \op{G}\bm{(}\op{V}_{\rm free}(-\alpha, \beta, -),\,q\bm{)}.
\end{equation}
If $\beta=0$, the global evolution operator is
\begin{equation}\label{eq:app_final_factorization_Vplus}
   \op{U} \bm{(} \op{V}_{\rm free}(\alpha,0,+) \bm{)}
   =e^{-i\op{H}_2} \,
   e^{i\op{H}_1(-\alpha,\, 1)} \, e^{-i\op{H}_2} \,
   e^{-i\op{H}_1(-\alpha,\, 1)} \, 
   e^{-i\op{H}_2} \,
   e^{i\op{H}_1(-\alpha,\,0)} \,
   e^{-i\op{H}_2} \, e^{-i\op{H}_1(-\alpha,\,0)}\,.
\end{equation}
If $\beta\neq0$, we can apply the same construction, together with the basis rotation~\eqref{eq:b_replacements}. This concludes our first proof of the integrability of $\op{U}\bm{(} \op{V}_{\rm free}(\alpha,\, \beta,\,\pm) \bm{)}$.

The JW transformation employed here also allows us to construct free-fermionic QCA built from neighborhood gates whose $\op{V}$s target $\geq 2$ qubits each. We outline this more general construction in App.~\ref{sec:larger_gates}.

\subsection{Mapping of certain Goldilocks QCA to free fermions from the six-vertex model}
\label{sec:mapping_six_v_proof}

Our second proof of integrability follows a strategy based on a mapping from the six-vertex model. This proof is more technical, so we sketch the main ideas here. Appendix~\ref{sec:6V} contains details.

The six-vertex model explains why H$_2$O-based ice has residual entropy at zero temperature~\cite{pauling1935structure}. Each molecule's oxygen can participate in one of six configurations with its four neighbor-molecules' hydrogens while obeying the ice condition, a constraint introduced below; hence the name \emph{six-vertex}~\cite{pauling1935structure,lieb1967residual}. The model and its generalizations relate to other problems in statistical mechanics and quantum many-body physics~\cite{kadanoff1971some, wu1969exact, baxter2008exactly}. Importantly, the model is exactly solvable~\cite{lieb1967residual,baxter1970exact,baxter2008exactly}.

The six-vertex model involves a square lattice~\cite{baxter2008exactly}. Each edge is empty or occupied---or, equivalently, carries an orientation: downward or upward. The net flux flowing into each vertex vanishes, according to the \emph{ice condition}. Six configurations satisfy this condition (Fig.~\ref{fig:six_vertex}).
\begin{figure}[h]
\begin{tikzpicture}
\draw (-0.5,-0.5) -- (0.5,0.5);
\draw (0.5,-0.5) -- (-0.5,0.5);
\node at (0,-1) {$a_1$};

\begin{scope}[shift={(2,0)}]
\draw[blue,line width=2] (-0.5,-0.5) -- (0.5,0.5);
\draw[blue,line width=2] (0.5,-0.5) -- (-0.5,0.5);
\node at (0,-1) {$a_2$};
\end{scope}

\begin{scope}[shift={(4,0)}]
\draw[blue,line width=2] (-0.5,-0.5) -- (0.5,0.5);
\draw (0.5,-0.5) -- (-0.5,0.5);
\node at (0,-1) {$b_1$};
\end{scope}

\begin{scope}[shift={(6,0)}]
\draw (-0.5,-0.5) -- (0.5,0.5);
\draw[blue,line width=2] (0.5,-0.5) -- (-0.5,0.5);
\node at (0,-1) {$b_2$};
\end{scope}

\begin{scope}[shift={(8,0)}]
\draw[blue,line width=2] (-0.5,-0.5) -- (0,0)  -- (-0.5,0.5);
\draw (0.5,-0.5) -- (0,0) -- (0.5,0.5);
\node at (0,-1) {$c_1$};
\end{scope}

\begin{scope}[shift={(10,0)}]
\draw (-0.5,-0.5) -- (0,0)  -- (-0.5,0.5);
\draw[blue,line width=2] (0.5,-0.5) -- (0,0) -- (0.5,0.5);
\node at (0,-1) {$c_2$};
\end{scope}
\end{tikzpicture}
\caption{\emph{Six-vertex model's allowed vertices.} 
Thick, blue lines represent downward-oriented edges (equivalently, occupied edges). Thin, black lines represent upward-oriented edges (equivalently, empty edges). The net flux at each vertex vanishes, according to the ice condition. The allowed vertex configurations have classical Boltzmann weights (or quantum transition amplitudes) $a_i$, $b_i$, and $c_i$, wherein $i=1,2$.}
\label{fig:six_vertex}
\end{figure}

In the original, classical model, the vertex types correspond to the Boltzmann weights $a_1,\, a_2,\, b_1,\, b_2,\, c_1,$ and $c_2$. Two-dimensional crystals of ice-conditioned vertices are Yang–Baxter-integrable~\cite{lieb1967residual,baxter2008exactly} and, under the condition $a_1a_2 +b_1b_2 = c_1c_2$, noninteracting~\cite{baxter1970exact,baxter2008exactly}. Surprisingly, this model (and related ones) can describe quantum dynamics~\cite{vanicat2018integrable,Miao2023integrablequantum}. The vertex weights become complex-valued transition amplitudes. One spatial dimension becomes a discrete-time dimension. Ice-conditioned vertices encode certain QCA-neighborhood constraints. To prove this fact, we find transition amplitudes that encode the Goldilocks constraint and unitarity, (App.~\ref{sec:6V}). The resulting instance of the six-vertex model is the Goldilocks QCA with the update unitary $\op{V}_{\rm free}$ [Eq.~\eqref{eq:V_free}] and is noninteracting.

Our approach contrasts with recent work. For example, statistical-physics methods enable classical simulations of certain random quantum circuits \cite{zhou2019emergent,fisher2023random,napp2022efficient}. Additionally, Refs.~\cite{prosen2021many,prosen2021reversible,pozsgay2021integrable} demonstrate certain QCAs' Yang–Baxter integrability by constructing a transfer matrix.  In contrast, (i) our system is not a random circuit, and (ii) we find a mapping to our system from a known free-fermionic model. 

\section{Identifying local charges}
\label{sec:identify_charges}

In this subsection, we exhibit local charges conserved by the discrete dynamics entailed in~\eqref{eq:V_free}. Certain Goldilocks QCAs' Yang–Baxter integrability implies the existence of extensively many mutually commuting charges. In principle, we could construct these charges by leveraging the six-vertex model's known transfer matrix~\cite{baxter2008exactly}. Alternatively, we could calculate charges via free-fermionic methods~\cite{fagotti2013reduced,fagotti2014conservation}. Instead, we pursue an approach applicable to all $\op{V}$, not only $\op{V}_{\rm free}$: numerically searching for all charges whose supports are limited.

We identify charges by adapting the numerical algorithm of Ref.~\cite{prosen2007chaos}. In principle, the algorithm returns all local charges of a given support size $n$ by explicitly checking for conservation of all possible such charges. For this reason, the less local a charge, the more computation time it requires. By construction, the algorithm cannot detect quasilocal charges whose support is not an integer number of sites, but a decaying function. Setting $\op{A}^{(i)} \in \{\Id, \op{\sigma}^{x}, \op{\sigma}^{y}, \op{\sigma}^{z}\}$, a Pauli-operator product with support size $\leq n$ is $\op{A}^{(1)}_j \op{A}^{(2)}_{j+1} \ldots \op{A}^{(n)}_{j+n}$. 
%There are $(n+3)! / (3! n!)$ possible products. 
Notate the sum of a Pauli-operator product over all sites as
\begin{align}\label{eq:density_sum}
    [\op{A}^{(1)} \op{A}^{(2)} \ldots \op{A}^{(n)}] \coloneqq \sum_{j=0}^{L-1} \op{A}^{(1)}_{j} \op{A}^{(2)}_{j+1}\ldots \op{A}^{(n)}_{j+n}
\end{align}
and a sum over all even-indexed sites by
\begin{equation}\label{eq:density_even_sum}
    [[\op{A}^{(1)} \op{A}^{(2)} \ldots \op{A}^{(n)}]] \coloneqq \sum_{j=0}^{L/2-1} \op{A}^{(1)}_{2j} \op{A}^{(2)}_{2j+1}\ldots \op{A}^{(n)}_{2j+n}\,.
\end{equation}
Our algorithm searches for linear combinations of sums \eqref{eq:density_sum} or \eqref{eq:density_even_sum} that are conserved during one QCA timestep at a small system size.

We focus first on the single-site gates $\op{V}_{\rm free}(\alpha, 0, -)$, generalizing later. We find that $\op{U}{\bm (}\op{V}_{\rm free}(\alpha, 0, -)\bm{)}$ conserves 13 linearly independent charges $\op{Q}_{i=1,2,\ldots, 13}$ of support sizes $\leq 5$. We identified these charges by searching symbolically in Mathematica.\footnote{We identify certain equipment to specify the computational procedure adequately. Such identification is not intended to imply recommendation or endorsement of any product or service by NIST; nor is it intended to imply that the software identified is necessarily the best available for the purpose.}
They have the forms
\begin{subequations} \label{eq:charges}
\begin{align}
& \op{Q}_1 = [\op{\sigma}^z \op{\sigma}^z],\\
& \op{Q}_2 =\tan \alpha ~ [\op{\sigma}^x \op{\sigma}^x] + [\op{\sigma}^x \op{\sigma}^z] +[\op{\sigma}^z \op{\sigma}^x],\\
& \op{Q}_3 = [\op{\sigma}^x\op{\sigma}^y\op{\sigma}^z] - [\op{\sigma}^z\op{\sigma}^y\op{\sigma}^x],\\
& \op{Q}_4 = [\op{\sigma}^y] + [\op{\sigma}^z\op{\sigma}^y\op{\sigma}^z] + \tan \alpha ~ [\op{\sigma}^x\op{\sigma}^y\op{\sigma}^z],\\
& \op{Q}_5 = [\op{\sigma}^x\op{\sigma}^x] + [\op{\sigma}^z\op{\sigma}^y\op{\sigma}^y\op{\sigma}^z],\\
& \op{Q}_6 = \tan \alpha ~ \left ( [\op{\sigma}^x\op{\sigma}^y\op{\sigma}^y\op{\sigma}^x] - [\op{\sigma}^z\op{\sigma}^y\op{\sigma}^y\op{\sigma}^z] \right )+ [\op{\sigma}^x\op{\sigma}^y\op{\sigma}^y\op{\sigma}^z] + [\op{\sigma}^z\op{\sigma}^y\op{\sigma}^y\op{\sigma}^x], \\
& \op{Q}_7 = [\op{\sigma}^x\op{\sigma}^y\op{\sigma}^x] - [\op{\sigma}^z\op{\sigma}^y\op{\sigma}^y\op{\sigma}^y\op{\sigma}^z] 
         - \tan \alpha ~ \left ([\op{\sigma}^x\op{\sigma}^y\op{\sigma}^z] + [\op{\sigma}^x\op{\sigma}^y\op{\sigma}^y\op{\sigma}^y\op{\sigma}^z] \right ),\\
& \op{Q}_8 = [\op{\sigma}^x\op{\sigma}^y\op{\sigma}^y\op{\sigma}^y\op{\sigma}^z] - [\op{\sigma}^z\op{\sigma}^y\op{\sigma}^y\op{\sigma}^y\op{\sigma}^x],\\
& \op{Q}_9 = \tan \alpha ~ \left ( [[\op{\sigma}^x\op{\sigma}^x]] -[[\op{\sigma}^z \op{\sigma}^z]] \right )+ [[\op{\sigma}^x \op{\sigma}^z]] + [[\op{\sigma}^z \op{\sigma}^x]], \\
& \op{Q}_{10} = [[\op{\sigma}^x \op{\sigma}^y \op{\sigma}^z]] - [[\op{\sigma}^z\op{\sigma}^y\op{\sigma}^x]], \\
& \op{Q}_{11} = [[\op{\sigma}^y]] + [[\op{\sigma}^z\op{\sigma}^y\op{\sigma}^z]] +\tan \alpha ~ [[\op{\sigma}^x \op{\sigma}^y \op{\sigma}^z]], \\
& \op{Q}_{12} = [[\op{\sigma}^x \op{\sigma}^x]] + [[\op{\sigma}^z \op{\sigma}^y \op{\sigma}^y \op{\sigma}^z]], \quad \text{and} \\
& \op{Q}_{13} = \tan \alpha ~ \left ( [[\op{\sigma}^x\op{\sigma}^y\op{\sigma}^y \op{\sigma}^x]] - 
[[\op{\sigma}^z\op{\sigma}^y\op{\sigma}^y \op{\sigma}^z]] \right )+ [[\op{\sigma}^x\op{\sigma}^y\op{\sigma}^y \op{\sigma}^z]] + [[\op{\sigma}^z\op{\sigma}^y\op{\sigma}^y \op{\sigma}^x]].
\end{align}
\end{subequations}
The charges are manifestly quadratic in the fermionic operators, by our JW transformation~\eqref{eq:fermionic}. The charges depend on Pauli-operator products $\sigma^{\gamma_1}_j\sigma^{y}_{j+1}\ldots \sigma^y_{j+q}\sigma^{\gamma_2}_{j+q+1}$---quadratic functions of fermionic operators associated with sites $j$ and $j+q+1$.

$\op{Q}_1$ counts domain walls. Every update $\op{V}$ conserves $\op{Q}_1$, even if we choose $\op{V}$ randomly at every site and circuit layer. Being diagonal relative to the computational basis, $\op{Q}_1$ is useful for postselective error mitigation on quantum hardware \cite{bonet2018low,jones2022small, rotello2023automated}.

The charges~\eqref{eq:charges} do not all mutually commute:
$[\op{Q}_i, \op{Q}_{i'}]$ can be nonzero, if $i' \neq i$. Our observation of noncommuting charges is consistent with Ref.~\cite{fagotti2014conservation}, since the dynamics map to a free-fermionic model that breaks single-site-translation symmetry.  Exploring thermodynamic consequences of noncommuting charges is a recent endeavor \cite{majidy2023noncommuting, Lostaglio_2017,guryanova2016thermodynamics,yunger2016microcanonical,ares2023lack,borsi2023matrix}; Goldilocks QCA provide an additional example of non-Abelian integrability~\cite{fukai2020noncommutative,corps2023general,potter2016symmetry,de2021stability}.

We generalize from $\op{U}\bm{(}\op{V}_{\rm free}(\alpha, 0, -)\bm{)}$ in two ways.  First, consider changing the single-qubit gate's final argument to $+$: $\op{V}_{\rm free}(\alpha, 0, +)$. The algorithm of Ref.~\cite{prosen2007chaos} enables symbolic and numerical searches for the charges. We applied the symbolic search method to generic $\alpha$ values, but the calculation did not terminate. Therefore, we turned to numerical search. Using it, we numerically found $\geq 9$ charges with support sizes $\leq 5$, for every $\alpha$ value sampled.  
According to~\cite{pozsgay2021integrable}, at $\alpha = \pi/2$, $\op{U}\bm{(}\op{V}_{\rm free}(\pi/2,0,+)=\op{\sigma}^x\bm{)}$ conserves $\op{Q}_1$ (maps to free domain-wall evolution) and conserves 
$\op{Q} = [[\op{I}\op{\sigma}^x\op{\sigma}^x\op{I}]] + [[\op{\sigma}^z\op{\sigma}^y\op{\sigma}^y\op{\sigma}^z]]+[[\op{\sigma}^z\op{\sigma}^z\op{\sigma}^z\op{\sigma}^z]]$.\footnote{Reference~\cite{pozsgay2021integrable} discusses neither mappings to free fermions nor the gate~\eqref{eq:V_free} with general parameter values.}
In addition to proving this QCA's free-fermionic nature, we numerically found 24 charges with support sizes $\leq 5$. Furthermore, we find that $\op{U}\bm{(}\op{\sigma}^x\bm{)}$ conserves each of the three terms in $\op{Q}$. Second, consider generalizing the second argument to $\beta\in[0,\,2\pi)$. The charges' forms follow from the $\beta=0$ charges and the change of basis~\eqref{eq:b_replacements}. 

Finally, we illustrate generic Goldilocks QCA. We construct their single-site update unitaries $\op{V}(a,b)$ by drawing the real and imaginary parts of $a$ and $b$ from independent standard normal distributions. Analyzing the representative example $\op{U}\bm{(}\op{V}(0.3 + 0.7i, -1.0 - 0.5 i)\bm{)}$, we find no charges (except $\op{Q}_1$) of support size $\leq 5$. This outcome, as well as evidence presented next, suggests that generic Goldilocks QCA are nonintegrable. 

\section{Simulating Goldilocks-QCA dynamics classically}
\label{sec:simulation_dynamics}

We now contrast integrable and generic Goldilocks QCAs' dynamics. As in the previous section, we construct generic Goldilocks QCA by choosing single-site update unitaries $\op{V}(a,b)$. We choose the real and imaginary parts of $a$ and $b$ from independent standard normal distributions. 

Our calculations proceed as follows.
Defining the tilted single-qubit state
$\ket{\theta, \phi} \coloneqq \cos (\theta/2) \ket{0}+ e^{i \phi} \sin (\theta/2) \ket{1}$,
we initialize the system in the tilted-ferromagnet state 
$\ket{\Psi_0(\theta, \phi) } \coloneqq \ket{\theta, \phi}^{\otimes L}$. 
We calculate time-dependent local expectation values
\begin{equation}\label{eq:correlations}
    \mathcal{F}(t;\gamma,k)
    \coloneqq \braket{\Psi_t|(\op{\sigma}^{\gamma})^{\otimes k} |\Psi_t}\,,
\end{equation} 
wherein $(\op{\sigma}^{\gamma})^{\otimes k}=\sigma^\gamma_0\otimes \ldots \otimes \sigma^\gamma_{k-1}\otimes \openone_{k} \ldots \openone_{L-1} \, .$

\begin{figure}
\includegraphics[scale=0.9]{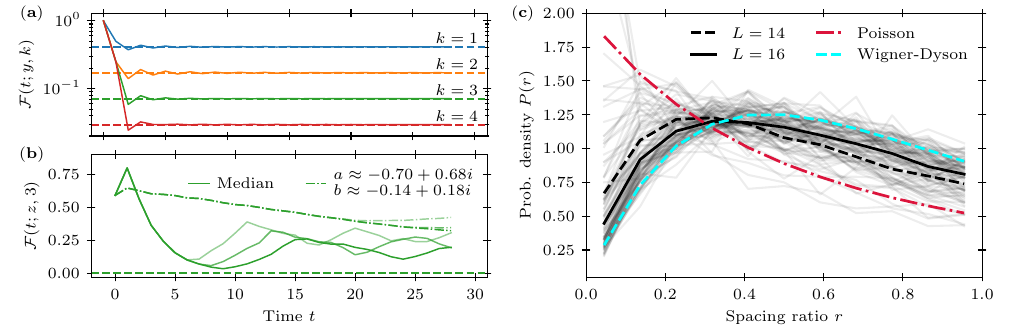}
\caption{\emph{Expectation-value dynamics and level statistics.} 
(a) Dynamics generated by integrable Goldilocks QCA $\op{U} \LParen \op{V}_{\rm free}(\pi/4, 0,-) \RParen$, for a system size $L=256$. The initial (Gaussian) state is the $y$-ferromagnet. 
(b) Median expectation value from 100 realizations of generic Goldilocks QCA $\op{U}{\bm (}\op{V}(a,b){\bm )}$ (solid curves) and one outlying realization [$96^{\rm th}$ percentile of $\mathcal{F}(9;z,3)$, dot-dashed curves]. Darker curves signal larger system sizes ($L=18, 22, 26$). The initial state is the tilted ferromagnet $\ket{\Psi_0(0.56, 3.7)}$. In (a)–(b), horizontal dashed curves mark the thermodynamic limits predicted by the appropriate (truncated) GGE.
(c) Distribution $P(r)$ over level-spacing ratios. The solid black curve shows the median distribution, calculated from 100 realizations of $\op{V}(a,b)$ at $L=16$, from the $(K{=}1, \, q_1{=}0)$ sector's 3,200 levels.
    The faint curves show individual realizations.
    The black dashed line shows the median distribution for the same 100 $\op{V}$ realizations but a smaller system ($L=14$) and the $(K{=}1, \, q_1{=}2)$ sector's 858 levels. The Wigner-Dyson (dashed, cyan) and Poisson (dot-dashed, red) distributions serve as references.
}
\label{fig:numerics}
\end{figure}

For the free-fermionic integrable Goldilocks QCA, we can calculate $\mathcal{F}(t;\gamma,k)$ classically, via the Gaussian formalism~\cite{bravyi2004lagrangian, surace2022fermionic}, given certain initial states. These are the states mapped, via the JW transformation, to Gaussian states. A covariance matrix specifies a Gaussian state, and Gaussian-state expectation values follow from Wick's theorem~\cite{bravyi2004lagrangian}. Hamiltonians quadratic in the fermionic creation and anhilation operators, such as Eqs. \eqref{eq:H1} and~\eqref{eq:H2}, map Gaussian states to Gaussian states. Therefore, instead of simulating a state vector in a $2^L$-dimensional Hilbert space, we need only evolve a $2L \times 2L$ covariance matrix (App.~\ref{sec:numerics}). The Gaussian states include the $y$-ferromagnet state, $\ket{\Psi_0(\pi/2,\pi/2)}=[(\ket{0}+i\ket{1})/\sqrt{2}]^{\otimes L}$, which we take as the initial state for our integrable Goldilocks QCA. The Gottesman–Knill theorem does not guarantee that classical computers can simulate this state's evolution: the initial state is not in the computational basis, and $\op{V}_{\rm free}$ involves arbitrary rotations about qubits' $x$- and $y$-axes \cite{aaronson2004improved}. Figure~\ref{fig:numerics}(a) illustrates our results. It shows expectation values $\mathcal{F}(t;y,k)$ of the experimentally realized Goldilocks QCA, whose $\op{V} = \op{V}_{\rm free}(\pi/4,0,-)$~\cite{jones2022small}. However, we simulate a system much larger than the experimental one, a system of size $L=256$. 

In contrast to free-fermionic QCA, generic QCA do not fit within the Gaussian formalism: they need not map Gaussian states to Gaussian states. Our numerics therefore involve brute-force representations of the evolution unitaries relative to the computational basis. We can achieve only small sizes and short times. Figure~\ref{fig:numerics}(b) illustrates the median (and one example outlier) of the expectation values $\mathcal{F}(t;z,3)$ calculated from 100 $\op{V}(a,b)$ realizations.  [In each realization, the same $\op{V}(a,b)$ acted at all sites and time steps.] We calculated $z$-correlation functions to facilitate analytic computations involving the GGE (App.~\ref{sec:Q1_GGE}). The median equilibrates towards zero as time and system size increase. As shown in the next section, this behavior is expected for every system that conserves only $Q_1$. There, we also physically interpret the observed outliers. We have also studied higher-order correlations, obtaining similar conclusions.

\subsection{Generalized Gibbs ensembles}
 The conserved charges implied by integrebility have consequences for the expectation values' time-averages in the large-system limit. In this \emph{thermodynamic limit}, expectation values are expected to be consistent with generalized Gibbs ensembles (GGEs)~\cite{rigol2008thermalization,vidmar2016generalized,essler2016quench}
\begin{equation} \label{eq:tGGE}
    \op{\rho} \coloneqq \frac{1}{Z} \exp \left( -\sum_i \op{Q}_i \mu_i \right )\, .
\end{equation}
The sum runs over the charges. The normalization factor $Z \coloneqq \Tr \bm{(} \exp ( -\sum_i \op{Q}_i \mu_i ) \bm{)}$. The $\mu_i$s denote generalized chemical potentials (Lagrange multipliers) fixed by the initial conditions: $\bra {\Psi_0 } \op{Q}_i \ket{\Psi_0} = \Tr (\op{\rho}\op{Q}_i)$. 
If one does not know all the charges' forms, one can truncate the GGE ~\cite{fagotti2013reduced,pozsgay2017generalized}---construct the approximate ensemble $\op{\rho}^{[n]}$ that includes the $M$ charges of support size $\leq n$.\footnote{If the system consists of free fermions, one can compute the full GGE predictions via the quench-action method~\cite{caux2013time,caux2016quench}. For the few-body expectation values we analyze, the simpler truncated GGEs produce sufficiently accurate predictions~\cite{fagotti2013reduced}.}
Truncated-GGE predictions approximate GGE predictions about observables with support sizes $<n$~\cite{fagotti2013reduced}. If all the charges commute pairwise, Eq.~\eqref{eq:tGGE} predicts time-averaged expectation values of a large system prepared in a microcanonical subspace---with well-defined values of the included charges~\cite{Landau_Statistical_Book}. If the charges fail to commute, Eq.~\eqref{eq:tGGE} predicts analogously for a system in an approximate microcanonical subspace---with fairly well-defined values of the included charges (since noncommuting charges cannot necessarily have well-defined values simultaneously)~\cite{yunger2016microcanonical,yungernoncommuting2020,kranzl2023experimental}. Our initial condition, being a product state, satisfies this requirement~\cite{yungernoncommuting2020}.

We computed truncated-GGE predictions $\Tr (\op{\rho}^{[n]}[\op{\sigma}^{\gamma}]^{\otimes k})$ for our free-fermionic and generic Goldilocks QCA. Figure~\ref{fig:numerics}(a)-(b) shows these results and comparisons with the exact time-evolved expectation values. The free-fermionic calculation involves the initial state $\ket{\Psi_0(\pi/2,
\,\pi/2)}$ and the 13 local charges~\eqref{eq:charges} with maximal support size $n=5$. The chemical potentials $\mu_i = 0$ (up to our method's numerical accuracy) for $i>M=8$.
The time-evolved expectation values rapidly equilibrate to the truncated-GGE predictions. As the observable-support size $k$ increases, the equilibrium value approaches zero (App.~\ref{sec:Q1_GGE}).

Generic Goldilocks QCA conserve only $\op{Q}_1$: The GGE in this case consists of $M=1$ charge with support size $n=2$. For an initial tilted-ferromagnet state with a polar angle $\theta$ and any azimuthal angle $\phi \in [0,\,2\pi)$, the chemical potential is $\mu_1 = {\rm arctanh}(\cos^2 \theta)$. We thus expect $\mathcal{F}(t;z,k)$ to equilibrate to $\Tr (\op{\rho}^{[2]}[\op{\sigma}^{z}]^{\otimes k}) = [\cos^k \theta + (-\cos \theta)^k]/2$. Appendix~\ref{sec:Q1_GGE} details the calculation of $\mu_1$ and $\Tr (\op{\rho}^{[2]}[\op{\sigma}^{z}]^{\otimes k})$. As described in the previous section, we have simulated $\mathcal{F}(t;z,3)$ over 100 realizations of $V(a,b)$. The median behaves as $\mathcal{F}(t,z,3) \to 0$, approaching the truncated-GGE prediction, as $L$ increases. Outliers equilibrate slowly in time and in $L$. Possible reasons include (i) the models' proximity to integrable points and (ii) conserved charges undetected by our method. Generic Goldilocks QCA appear nonintegrable, however, obeying predictions by a thermal state dependent on only one charge. The next section provides additional support for this conjecture. There, we show that generic Goldilocks QCAs' symmetry-resolved quasienergy spectra follow Wigner–Dyson statistics, a hallmark of nonintegrability.

\subsection{Level statistics}\label{sec:level_statistics}

(Quasi)energy-level statistics can evidence integrability~\cite{prosen2007chaos,giraud2022probing,prosen2021reversible}. Roughly speaking, we can distinguish two cases: (i) In the absence of conservation laws, levels repel each other.
% level repulsion is expected, leading to a 
A Wigner-Dyson probability distribution captures the spectrum's low probability of containing small gaps. (ii) If a system is integrable, conservation laws imply (quasi)energy degeneracies and level crossings. Poisson statistics model the spectrum's high probability of containing small spectral gaps. One must compute level statistics from energy levels in the same symmetry sector. If one fails to do so, unresolved symmetries cause degeneracies that confound the interpretation of the level statistics. In this section, we outline our procedure for resolving the quasimomentum and $Q_1$ symmetries of generic Goldilocks QCA. We then detail the statistical analysis of symmetry-resolved level-spacing ratios that arise in generic Goldilocks QCA. Most generic Goldilocks QCA are consistent with nonintegrability, we demonstrate.

We computed symmetry-resolved spectral statistics as follows. Denote by $\op{\Pi}$ the operator that shifts the system by one site. (The shift's direction does not impact the rest of our argument.) The Goldilocks dynamics conserve the \emph{two}-site-shift operator $\op{\Pi}^2$, due to the circuit's brickwork structure: $[\op{U}(\op{V}), \, \op{\Pi}^2] = 0$, for all $\op{V}$. 
Hence this the two-site-shift operator is a conserved charge. We can regard it as a conserved quasimomentum.
Not only $\op{U}$, but also the already known charge $\op{Q}_1$, commutes with $\op{\Pi}^2$: 
$[\op{\Pi}^2, \, \op{Q}_1]=0$. 
Therefore, the two charges share an eigenbasis 
$\{ \ket{\psi_j} \}$. 
Denote the $\op{Q}_1$ eigenvalues by 
$q_1 = -L,-L+4,\ldots,L$; and the $\op{\Pi}^2$ eigenvalues, by 
$e^{4 \pi i K /L}$, wherein
$K\in 0,1,\ldots,L/2$. Consider the joint eigenspace, shared by the two charges, labeled by some quantum numbers $q_1$ and $K$. 
Denote the subspace's dimensionality by $N$.
Define the projector onto this subspace as
$\op{R} \coloneqq \sum_{j=0}^{N-1} \ket{\psi_j} \bra{ \psi_j}$.
(We suppress $q_1$ and $K$ labels for notational convenience.)
From this projector, we construct the unitary operator
$\op{B} =\op{R}\op{\Pi} \op{G}(\op{V}, 0)\op{R}^{\dagger}$.

The eigenvalues of $\op{B}$ hold information about the integrability of Goldilocks QCA updated via unitary $\op{U}$. Following \cite{prosen2021reversible}, we compute the eigenvalues of $\op{B}$, denoted $e^{i \phi_{\ell}}$. 
The quasienergies $\phi_{\ell}$ are indexed by $\ell=1,2\ldots N$. We order the quasienergies such that $\phi_1 \leq  \phi_2 \leq \ldots \leq \phi_{N}$. Then, we compute the spacings $s_{\ell} \coloneqq \phi_{\ell+1} - \phi_{\ell}$. The smallest ratio of adjacent spacings is $r_{\ell} \coloneqq \min\{ s_{\ell},\, s_{\ell-1} \} / \max \{ s_{\ell'},\, s_{\ell'-1} \}$. $P(r)$ denotes the distribution, calculated across the quasienergy spectrum, over these ratios. $P(r)$ is more convenient than the distribution over the spacings themselves, which require spectral-unfolding analyses~\cite{oganesyan2007localization,atas2013distribution,giraud2022probing}. \emph{Poisson} statistics can signal integrability~\cite{berry1977level}:
$P_{\rm P}(r) \coloneqq 2/(1+r)^2 \, .$
In contrast, \emph{Wigner-Dyson} statistics can signal nonintegrability~\cite{bohigas1984characterizing}: 
$P_{\rm WD}(r) \coloneqq (27/4) (r+r^2)/(1+r+r^2)^{5/2} \, .$ 

Figure~\ref{fig:numerics}(c) shows the level statistics of typical single-site update operators $\op{V}(a, b)$. Different realizations evidence Wigner-Dyson-like, Poisson-like, and intermediate statistics. The median distribution is closest to Wigner-Dyson. As the system size $L$ grows, so does the fraction that exhibits Wigner-Dyson-like statistics. Therefore, most Goldilocks QCA appear nonintegrable.

\section{Outlook}
\label{sec:outlook}

We have proven that certain Goldilocks QCA exhibit free-fermionic integrability. This claim rests on two proofs, one involving a Jordan--Wigner transformation and one, a mapping from the six-vertex model. We have demonstrated implications of this integrability: classical simulability, many noncommuting local charges, and rapid equilibration to truncated-GGE predictions. Generic Goldilocks QCA are predominantly consistent with nonintegrability, as evidenced by (i) the discovery of only one local charge  (ii) thermalization to single-charge-GGE predictions and (iii) Wigner-Dyson-like level statistics. 

Some Goldilocks QCA exhibit anomalous behaviors that invite future research: intermediate level statistics and thermalization that is less complete, at finite $L$, than one might expect. Noncommuting charges may enable such behaviors~\cite{majidy2023noncommuting}, as may near-integrability. Do anomalous Goldilcoks QCA lie close to integrable parameter points, conserve hidden noncommuting charges, or approximately conserve noncommuting charges? Noncommuting-charge studies have focused on exact conservation~\cite{majidy2023noncommuting}. Extensions to approximate conservation could uncover further quantum thermodynamics~\cite{25_Campbell_Roadmap}, and Goldilocks QCA might provide a toy model.

Our integrability proofs raise more questions. Our six-vertex map works for many QCA but, applied to Goldilocks QCA, reveals noninteracting integrability. Could other six-vertex QCA exhibit interacting integrability~\cite{wintermantel2020unitary,sa2021integrable}? Can one classify all the QCA that map to free fermions? As shown in App.~\ref{sec:larger_gates}, our free-fermionic gates generalize to ones built from neighborhood gates whose $\op{V}$s target $\geq 2$ qubits each. More generally, could QCA feature free fermions in disguise like those in Ref.~\cite{fendley2019free}? The recent papers~\cite{fukai2025quantum,szasz2025construction} suggests so. The above questions' answers may identify settings for quantum advantages and provide data for benchmarking.
\newline

\emph{Acknowledgements.}---We acknowledge useful conversations with Ehud Altman, Mina Fasihi, Matthew Jones, Norman Margolus, and Andrew Potter.  This work was performed in part with support by the U.S. National Science Foundation under grants PHY-2210566, PHY-2515059, and OMA-2120757, and Slovenian Research and Innovation Agency (ARIS) under grants P1-0402, N1-0334, N1-0219 (TP).
\\

\emph{Data availability statement}.-- The data that support the findings of this study are openly available at the following URL/DOI: 10.5281/zenodo.18180826.

\appendix

\section{Mapping from the six-vertex model}\label{sec:6V}

In this section, we detail the mapping from the six-vertex model to certain Goldilocks QCA. We start by recalling elementary facts about the model. 
Then, we show that the six-vertex model maps to a set of QCA neighborhood gates. This set of gates overlaps partially with the set of QCA neighborhood gates $\op{u}_j(\op{V})$ focused on in this paper [Eq.~\eqref{eq:uj}].
Finally, we identify the Goldilocks QCA mapped to by points in the six-vertex model's parameter space. At these points, we show, the six-vertex model is free-fermionic.

\subsection{Six-vertex model} \label{sec_6V_backgrnd}

The six-vertex model~\cite{baxter2008exactly} is defined on a lattice whose degrees of freedom are the edges. Each edge is empty (thin, black) or occupied (thick, blue)---or, equivalently, carries an orientation: upwards for empty edges and downwards for occupied edges. The net flux flowing into each vertex must vanish. According to this \emph{ice condition}, the allowed vertices are
\begin{equation}
\begin{tikzpicture}
\draw (-0.5,-0.5) -- (0.5,0.5);
\draw (0.5,-0.5) -- (-0.5,0.5);
\node at (0,-1) {$a_1$};

\begin{scope}[shift={(2,0)}]
\draw[blue,line width=2] (-0.5,-0.5) -- (0.5,0.5);
\draw[blue,line width=2] (0.5,-0.5) -- (-0.5,0.5);
\node at (0,-1) {$a_2$};
\end{scope}

\begin{scope}[shift={(4,0)}]
\draw[blue,line width=2] (-0.5,-0.5) -- (0.5,0.5);
\draw (0.5,-0.5) -- (-0.5,0.5);
\node at (0,-1) {$b_1$};
\end{scope}

\begin{scope}[shift={(6,0)}]
\draw (-0.5,-0.5) -- (0.5,0.5);
\draw[blue,line width=2] (0.5,-0.5) -- (-0.5,0.5);
\node at (0,-1) {$b_2$};
\end{scope}

\begin{scope}[shift={(8,0)}]
\draw[blue,line width=2] (-0.5,-0.5) -- (0,0)  -- (-0.5,0.5);
\draw (0.5,-0.5) -- (0,0) -- (0.5,0.5);
\node at (0,-1) {$c_1$};
\end{scope}

\begin{scope}[shift={(10,0)}]
\draw (-0.5,-0.5) -- (0,0)  -- (-0.5,0.5);
\draw[blue,line width=2] (0.5,-0.5) -- (0,0) -- (0.5,0.5);
\node at (0,-1) {$c_2$};
\end{scope}
\end{tikzpicture}
\end{equation}

As originally formulated, the six-vertex model is a classical statistical-mechanics model in two spatial dimensions. Each vertex is assigned a positive Boltzmann weight---$a_1$, $a_2$, $b_1$, $b_2$, $c_1$, or $c_2$---as indicated above. 
A lattice configuration has a Boltzmann weight equal to the product of its vertices' weights. The model's partition function is the sum over the configurations' lattice weights. The six-vertex model's parameter space has free-fermionic points specified by the condition $a_1a_2+b_1b_2=c_1c_2$~\cite{baxter1970exact,baxter2008exactly}.

This vertex model, whose degrees of freedom are the edges, is equivalent to a model whose degrees of freedom are spins~\cite{wu1969exact,kadanoff1971some,baxter2008exactly,bazhanov2023ising}. Let a classical spin live on each face of each vertex, such that the vertex becomes the center of a plaquette. We represent an upward-pointing spin with a 0 and a downward-pointing spin with a 1. The vertex model is equivalent to this spin model under the following rule: empty (filled) lattice edges separate neighboring spins that are aligned (antialigned).
Two plaquette configurations are consistent with each of the six vertex configurations\footnote{
One might hesitate to label as equivalent two models that have different numbers of allowed configurations. These different numbers, however, merely distinguish one model's partition function from the other model's by a factor of two~\cite{wu1969exact}. Hence the term \emph{equivalent} appears in~\cite{wu1969exact}\label{foot_equiv}}.
These ice-conditioned plaquettes, and the corresponding weights, are
\begin{equation}
\begin{tikzpicture}
\draw (-0.5,-0.5) -- (0.5,0.5);
\draw (0.5,-0.5) -- (-0.5,0.5);
\node at (-0.5,0) {$0$};
\node at (0.5,0) {$0$};
\node at (0,0.5) {$0$};
\node at (0,-0.5) {$0$};

\begin{scope}[shift={(2,0)}]
\draw[blue,line width=2] (-0.5,-0.5) -- (0.5,0.5);
\draw[blue,line width=2] (0.5,-0.5) -- (-0.5,0.5);
\node at (-0.5,0) {$0$};
\node at (0.5,0) {$0$};
\node at (0,0.5) {$1$};
\node at (0,-0.5) {$1$};
\end{scope}

\begin{scope}[shift={(4,0)}]
\draw (-0.5,-0.5) -- (0.5,0.5);
\draw[blue,line width=2] (0.5,-0.5) -- (-0.5,0.5);
\node at (-0.5,0) {$0$};
\node at (0.5,0) {$1$};
\node at (0,0.5) {$1$};
\node at (0,-0.5) {$0$};
\end{scope}

\begin{scope}[shift={(6,0)}]
\draw[blue,line width=2] (-0.5,-0.5) -- (0.5,0.5);
\draw (0.5,-0.5) -- (-0.5,0.5);
\node at (-0.5,0) {$0$};
\node at (0.5,0) {$1$};
\node at (0,0.5) {$0$};
\node at (0,-0.5) {$1$};
\end{scope}

\begin{scope}[shift={(8,0)}]
\draw[blue,line width=2] (-0.5,-0.5) -- (0,0)  -- (-0.5,0.5);
\draw (0.5,-0.5) -- (0,0) -- (0.5,0.5);
\node at (-0.5,0) {$1$};
\node at (0.5,0) {$0$};
\node at (0,0.5) {$0$};
\node at (0,-0.5) {$0$};
\end{scope}

\begin{scope}[shift={(10,0)}]
\draw (-0.5,-0.5) -- (0,0)  -- (-0.5,0.5);
\draw[blue,line width=2] (0.5,-0.5) -- (0,0) -- (0.5,0.5);
\node at (-0.5,0) {$0$};
\node at (0.5,0) {$1$};
\node at (0,0.5) {$0$};
\node at (0,-0.5) {$0$};
\end{scope}

\begin{scope}[shift={(0,-1.5)}]
\draw (-0.5,-0.5) -- (0.5,0.5);
\draw (0.5,-0.5) -- (-0.5,0.5);
\node at (-0.5,0) {$1$};
\node at (0.5,0) {$1$};
\node at (0,0.5) {$1$};
\node at (0,-1) {$a_1$};
\node at (0,-0.5) {$1$};

\begin{scope}[shift={(2,0)}]
\draw[blue,line width=2] (-0.5,-0.5) -- (0.5,0.5);
\draw[blue,line width=2] (0.5,-0.5) -- (-0.5,0.5);
\node at (-0.5,0) {$1$};
\node at (0.5,0) {$1$};
\node at (0,0.5) {$0$};
\node at (0,-0.5) {$0$};
\node at (0,-1) {$a_2$};
\end{scope}

\begin{scope}[shift={(4,0)}]
\draw (-0.5,-0.5) -- (0.5,0.5);
\draw[blue,line width=2] (0.5,-0.5) -- (-0.5,0.5);
\node at (-0.5,0) {$1$};
\node at (0.5,0) {$0$};
\node at (0,0.5) {$0$};
\node at (0,-0.5) {$1$};
\node at (0,-1) {$b_1$};
\end{scope}

\begin{scope}[shift={(6,0)}]
\draw[blue,line width=2] (-0.5,-0.5) -- (0.5,0.5);
\draw (0.5,-0.5) -- (-0.5,0.5);
\node at (-0.5,0) {$1$};
\node at (0.5,0) {$0$};
\node at (0,0.5) {$1$};
\node at (0,-0.5) {$0$};
\node at (0,-1) {$b_2$};
\end{scope}

\begin{scope}[shift={(8,0)}]
\draw[blue,line width=2] (-0.5,-0.5) -- (0,0)  -- (-0.5,0.5);
\draw (0.5,-0.5) -- (0,0) -- (0.5,0.5);
\node at (-0.5,0) {$0$};
\node at (0.5,0) {$1$};
\node at (0,0.5) {$1$};
\node at (0,-0.5) {$1$};
\node at (0,-1) {$c_1$};
\end{scope}

\begin{scope}[shift={(10,0)}]
\draw (-0.5,-0.5) -- (0,0)  -- (-0.5,0.5);
\draw[blue,line width=2] (0.5,-0.5) -- (0,0) -- (0.5,0.5);
\node at (-0.5,0) {$1$};
\node at (0.5,0) {$0$};
\node at (0,0.5) {$1$};
\node at (0,-0.5) {$1$};
\node at (0,-1) {$c_2$};
\end{scope}
\end{scope}
\end{tikzpicture}
\label{weights6Vface_general}
\end{equation}

Instead of representing a classical spin system in two spatial dimensions, the six-vertex model can represent a one-dimensional quantum system evolving in discrete time~\cite{prosen2021many}. The classical lattice's $y$-axis can be reinterpreted as the quantum system's time axis, which runs from bottom to top. Consecutive lattice rows represent the quantum system's state at consecutive time steps. Similarly to in the classical spin model, a qubit lives on each face of each vertex. Each qubit's Hilbert space has a computational basis $\{ \ket{0}, \ket{1} \}$. If a qubit is in $\ket{0}$ ($\ket{1}$), we say that the spin is pointing upward (downward). Within a row, the southward spins specify the system's state at some time, and the northward spins specify the system's state at the next time step. Enforcing the ice condition constrains the qubits' evolution. 
Specifying the evolution, we specify, for each qubit, the complex amplitudes for the transitions $(0 \to 0)$, $(0 \to 1)$, $(1 \to 0)$, and $(1 \to 1)$.
The vertex weights serve as (un-normalized versions of) those transition amplitudes, assuming complex values.

\subsection{Quantum cellular automata mapped to by the six-vertex model}
 
The quantum lattice's weights can be encoded in QCA neighborhood gates~\cite{prosen2021many}. 
A vertex's southward spin represents an element of the target qubit's computational basis. The westward and eastward spins represent elements of the left-hand and right-hand neighboring control qubits' computational bases. The target qubit transitions from the southward to the northward spin state with the complex transition amplitude mentioned in the previous subsection. Below, we describe how to encode this amplitude in a QCA gate.

To identify the most general such gate, we scale the vertex weights. The allowed scalings are the ones that cancel in the lattice-configuration weight---the product of the vertices' weights. Index a plaquette's surrounding spins by $m,j,n,l\in \{0,1\}$, running clockwise from the westward face. Each vertex weight may be scaled by a factor $\omega_1^m \omega_2^j\omega_3^n \omega_4^l$, wherein $\omega_1 \omega_2 \omega_3 \omega_4 = 1$. 
The resulting weights assume the following forms: 
\begin{equation}
\begin{tikzpicture}
\draw (-0.5,-0.5) -- (0.5,0.5);
\draw (0.5,-0.5) -- (-0.5,0.5);
\node at (-0.5,0) {$0$};
\node at (0.5,0) {$0$};
\node at (0,0.5) {$0$};
\node at (0,-0.5) {$0$};
\node at (0,-1) {$a_1$};

\begin{scope}[shift={(2,0)}]
\draw[blue,line width=2] (-0.5,-0.5) -- (0.5,0.5);
\draw[blue,line width=2] (0.5,-0.5) -- (-0.5,0.5);
\node at (-0.5,0) {$0$};
\node at (0.5,0) {$0$};
\node at (0,0.5) {$1$};
\node at (0,-0.5) {$1$};
\node at (0,-1) {$a_2 \omega_2 \omega_4$};
\end{scope}

\begin{scope}[shift={(4,0)}]
\draw (-0.5,-0.5) -- (0.5,0.5);
\draw[blue,line width=2] (0.5,-0.5) -- (-0.5,0.5);
\node at (-0.5,0) {$0$};
\node at (0.5,0) {$1$};
\node at (0,0.5) {$1$};
\node at (0,-0.5) {$0$};
\node at (0,-1) {$b_1 \omega_2 \omega_3$};
\end{scope}

\begin{scope}[shift={(6,0)}]
\draw[blue,line width=2] (-0.5,-0.5) -- (0.5,0.5);
\draw (0.5,-0.5) -- (-0.5,0.5);
\node at (-0.5,0) {$0$};
\node at (0.5,0) {$1$};
\node at (0,0.5) {$0$};
\node at (0,-0.5) {$1$};
\node at (0,-1) {$b_2 \omega_3 \omega_4 $};
\end{scope}

\begin{scope}[shift={(8,0)}]
\draw[blue,line width=2] (-0.5,-0.5) -- (0,0)  -- (-0.5,0.5);
\draw (0.5,-0.5) -- (0,0) -- (0.5,0.5);
\node at (-0.5,0) {$1$};
\node at (0.5,0) {$0$};
\node at (0,0.5) {$0$};
\node at (0,-0.5) {$0$};
\node at (0,-1) {$c_1 \omega_1$};
\end{scope}

\begin{scope}[shift={(10,0)}]
\draw (-0.5,-0.5) -- (0,0)  -- (-0.5,0.5);
\draw[blue,line width=2] (0.5,-0.5) -- (0,0) -- (0.5,0.5);
\node at (-0.5,0) {$0$};
\node at (0.5,0) {$1$};
\node at (0,0.5) {$0$};
\node at (0,-0.5) {$0$};
\node at (0,-1) {$c_2 \omega_3$};
\end{scope}

\begin{scope}[shift={(0,-2.)}]
\draw (-0.5,-0.5) -- (0.5,0.5);
\draw (0.5,-0.5) -- (-0.5,0.5);
\node at (-0.5,0) {$1$};
\node at (0.5,0) {$1$};
\node at (0,0.5) {$1$};
\node at (0,-1) {$a_1$};
\node at (0,-0.5) {$1$};

\begin{scope}[shift={(2,0)}]
\draw[blue,line width=2] (-0.5,-0.5) -- (0.5,0.5);
\draw[blue,line width=2] (0.5,-0.5) -- (-0.5,0.5);
\node at (-0.5,0) {$1$};
\node at (0.5,0) {$1$};
\node at (0,0.5) {$0$};
\node at (0,-0.5) {$0$};
\node at (0,-1) {$a_2 \omega_1 \omega_3$};
\end{scope}

\begin{scope}[shift={(4,0)}]
\draw (-0.5,-0.5) -- (0.5,0.5);
\draw[blue,line width=2] (0.5,-0.5) -- (-0.5,0.5);
\node at (-0.5,0) {$1$};
\node at (0.5,0) {$0$};
\node at (0,0.5) {$0$};
\node at (0,-0.5) {$1$};
\node at (0,-1) {$b_1 \omega_4 \omega_1$};
\end{scope}

\begin{scope}[shift={(6,0)}]
\draw[blue,line width=2] (-0.5,-0.5) -- (0.5,0.5);
\draw (0.5,-0.5) -- (-0.5,0.5);
\node at (-0.5,0) {$1$};
\node at (0.5,0) {$0$};
\node at (0,0.5) {$1$};
\node at (0,-0.5) {$0$};
\node at (0,-1) {$b_2 \omega_1 \omega_2$};
\end{scope}

\begin{scope}[shift={(8,0)}]
\draw[blue,line width=2] (-0.5,-0.5) -- (0,0)  -- (-0.5,0.5);
\draw (0.5,-0.5) -- (0,0) -- (0.5,0.5);
\node at (-0.5,0) {$0$};
\node at (0.5,0) {$1$};
\node at (0,0.5) {$1$};
\node at (0,-0.5) {$1$};
\node at (0,-1) {$c_1 /\omega_1$};
\end{scope}

\begin{scope}[shift={(10,0)}]
\draw (-0.5,-0.5) -- (0,0)  -- (-0.5,0.5);
\draw[blue,line width=2] (0.5,-0.5) -- (0,0) -- (0.5,0.5);
\node at (-0.5,0) {$1$};
\node at (0.5,0) {$0$};
\node at (0,0.5) {$1$};
\node at (0,-0.5) {$1$};
\node at (0,-1) {$c_2 / \omega_3$};
\end{scope}
\end{scope}
\end{tikzpicture}
\label{eq:weights6Vface}
\end{equation}

We map these weights to QCA gates as follows. As noted in the previous subsection, each vertex weight can serve as a transition amplitude. Here, we interpret the weight as a target qubit's probability amplitude of transitioning from configuration $l$ to configuration $j$ if the westward neighbor is in configuration $m$ and the eastward neighbor is in $n$. We must specify four transition amplitudes: those for $(0 \to 0)$, $(0 \to 1)$, $(1 \to 0)$, and $(1 \to 1)$. We encapsulate these four transition amplitudes in four operators $\op{V}_{\rm 6V}(m,n)$. The operators are represented, relative to the computational basis, by
\begin{align}
\op{V}_{\rm 6V}(0,0) & \to
    \left(
\begin{array}{cc}
a_1 & 0 \\
0 & a_2 \omega_2 \omega_4
\end{array}
\right) , \label{eq:6vert_V00}\\
\op{V}_{\rm 6V}(0,1) & \to
    \left(
\begin{array}{cc}
c_2 \omega_3 & b_2 \omega_3 \omega_4 \\
b_1 \omega_2 \omega_3 & c_1 / \omega_1
\end{array}
\right) \, , \label{eq:6vert_V01}\\
\op{V}_{\rm 6V}(1,0) & \to 
    \left(
\begin{array}{cc}
c_1 \omega_1 & b_1 \omega_4 \omega_1 \\
b_2 \omega_1 \omega_2 & c_2 / \omega_3
\end{array}
\right)\, ,\quad \text{and}  \label{eq:6vert_V10} \\
\op{V}_{\rm 6V}(1,1) & \to 
    \left(
\begin{array}{cc}
a_2 \omega_1 \omega_3 & 0 \\
0 & a_1
\end{array}
\right) .  \label{eq:6vert_V11}
\end{align}
Using these single-site transition operators, we can represent a three-site neighborhood's evolution:
\begin{equation} \label{eq:uj_6V}
 \sum_{m,n=0}^{1}\ket{m}  \bra{m} \otimes \op{V}_{\rm 6V}(m,n) \otimes \ket{n}\bra{n} .
\end{equation}

This QCA operator resembles the QCA operator $\op{u}_j (\op{V})$ focused on throughout this paper [Eq.~\eqref{eq:uj}]. However, $\op{u}_j (\op{V})$ is more general in one way, whereas Eq.~\eqref{eq:uj_6V} is more general in another way. $\op{u}_j (\op{V})$ evolves the target as $\op{V}$ under some neighborhood configurations and as $\Id$ under other configurations. In contrast, under Eq.~\eqref{eq:uj_6V}, the target evolves under the more general $\op{V}_{\rm 6V}$. On the other hand, $\op{V}$ is not restricted to the form in Eqs.~\eqref{eq:6vert_V01}--\eqref{eq:6vert_V10}. 

In conclusion, the six-vertex model maps to QCA neighborhood gates of the form~\eqref{eq:uj_6V}. These gates form a set that overlaps partially with the set of QCA gates~\eqref{eq:uj} focused on in this paper. In the next subsection, we derive the forms of the Goldilocks QCA mapped to by the six-vertex model. 

\begin{figure}
\includegraphics[]{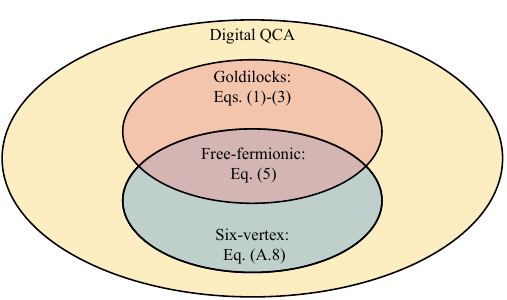}
\caption{ \emph{QCA Venn diagram.} 
  QCA compatible with both the six-vertex model and the Goldilocks constraint are free-fermionic.
}
\label{fig:venn_diagram}
\end{figure}

\subsection{Goldilocks quantum cellular automata mapped to by the six-vertex model}

We now specify the vertex weights under which the six-vertex model maps to Goldilocks QCA. As a reminder, a Goldilocks QCA effects the neighborhood gate $\op{u}_j$ of Eq.~\eqref{eq:uj}, subject to the constraint $f_{\rm G}(m, n)=m \oplus n$ [Eq.~\eqref{eq:goldilocks_rule}]. 
According to the constraint, the identity operator acts on a target qubit if the neighbors are in $\ket{00}$ or $\ket{11}$. This rule fixes $\op{V}_{\rm 6V}(0,0)=\op{V}_{\rm 6V}(1,1)=\Id$. Hence, by Eqs.~\eqref{eq:6vert_V00} and\eqref{eq:6vert_V11},
\begin{equation}\label{eq:identity_weights}
a_1=a_2\omega_2\omega_4 = a_2\omega_1\omega_3 =1. 
\end{equation}
If the neighboring spins are in $\ket{01}$ or $\ket{10}$, the target evolves under $\op{V}$. This rule fixes the remaining vertices' weights via the matrix equation $\op{V}_{\rm 6V}(0,1)=\op{V}_{\rm 6V}(1,0)=\op{V}$. That matrix equation consists of four components:
\begin{align}
\bra{0}\op{V}\ket{0}&=c_1\omega_1 =c_2 \omega_3,  \label{eq:V00_weights}\\
\bra{0}\op{V}\ket{1} &= b_2 \omega_3 \omega_4 = b_1 \omega_1 \omega_4, 
\label{eq:V01_weights}\\
\bra{1}\op{V}\ket{0} &= b_1 \omega_2 \omega_3 = b_2 \omega_1 \omega_2,  \; \: \text{and}
\label{eq:V10_weights} \\
\bra{1}\op{V}\ket{1} &=c_1 /\omega_1 = c_2 / \omega_3 \label{eq:V11_weights}.
\end{align}

To simplify notation, we use Eq.~\eqref{eq:identity_weights} to define $\epsilon_1 \equiv \omega_2 \omega_4 = \omega_1\omega_3 = \pm 1$. Equations~\eqref{eq:V00_weights} and~\eqref{eq:V11_weights} imply
$(\omega_3)^2 = ( \omega_1 )^2 \equiv \epsilon_2$.
Combining these two statements yields
$\epsilon_2 = \pm 1$. We can now express some of the original weights (some of the $a$, $b$, and $c$ numbers) in terms of $\epsilon_1$, $\epsilon_2$, and the other original weights: 
\begin{align}
  a_1 = \epsilon_1 a_2 = 1 \,, \qquad
  b_1 = \epsilon_1 \epsilon_2 b_2 \,, 
  \qquad \text{and} \qquad
  c_1 = \epsilon_1\epsilon_2 c_2 \,.
\end{align}
Hence the single-site Goldilocks gate realizable with the six-vertex model has the form
\begin{equation}
\op{V} = \left(
\begin{array}{cc}
c_1 \omega_1 &  b_1 \omega_1/\omega_2 \epsilon_1  \\ 
b_1 \omega_1 \omega_2 \epsilon_1 \epsilon_2 &   c_1 / \omega_1
\end{array}
\right)   \,.
\end{equation} 

The gate's unitarity fixes $|\omega_2|^2=1$, \: $|c_1|^2+|b_1|^2=1$, \: and \: $c_1^* b_1+c_1 b_1^*=0$. \: Up to a global phase factor in $\op{V}$, we parameterize 
$c_1 \omega_1= \cos\alpha$, \:
$b_1 \omega_1 \epsilon_1=i \sin\alpha$, \: and \:
$\omega_2 \epsilon_2 = -i e^{i\beta}$, \: wherein $\alpha,\beta \in \mathbb{R}$.
The single-site Goldilocks gate realizable with the six-vertex model acquires the form in the main text's Eq.~\eqref{eq:V_free}, which we reproduce here:
\begin{equation}
\op{V} = \left(
\begin{array}{cc}
\cos\alpha &  -\epsilon_2 e^{-i\beta}\sin\alpha \\ 
 e^{i\beta} \sin\alpha &  \epsilon_2 \cos\alpha
\end{array}
\right)   \,.
\end{equation} 
For any $\epsilon_1,\epsilon_2 = \pm 1$, the Goldilocks-constrained six-vertex model's weights satisfy 
\begin{equation} 
a_1 a_2 + b_1 b_2 - c_1 c_2  
= 
\epsilon_1 + \epsilon_1 \epsilon_2 ( \omega_1 i \sin\alpha )^2 - \epsilon_1 \epsilon_2 ( \omega_1 \cos \alpha)^2 
= \epsilon_1 (1- \sin^2 \alpha - \cos^2 \alpha ) = 0 \,,
\end{equation}  
i.e., the free-fermionic condition from two subsections ago.

We have shown that the six-vertex model, at free-fermionic points of its parameter space, maps to certain Goldilocks QCA (Fig.~\ref{fig:venn_diagram}). One can run this argument backward, to map these QCA to the free-fermionic six-vertex model. Hence the model and these QCA are equivalent, in the sense of the footnote near the end of the third paragraph in Sec.~\ref{sec_6V_backgrnd}.

\section{Details about numerical computations}\label{sec:numerics}

In this section, we further detail the efficient numerical simulation of the integrable Goldilocks QCA. As we have shown, the Goldilocks QCA~\eqref{eq:V_free} effect free-fermionic Floquet dynamics. The time-step operators $\op{U}(\op{V}_{\rm free})$ are Gaussian: they follow from multiplying exponentials of Hamiltonians that are quadratic in fermionic operators [Eq.~\eqref{eq:app_final_factorization}]. Consequently, the dynamics are efficiently simulated, if the system is initialized in a fermionic Gaussian state~\cite{bravyi2004lagrangian,surace2022fermionic}, as we now review. A state is Gaussian if it satisfies Wick's theorem~\cite{bravyi2004lagrangian,surace2022fermionic}.

For convenience of numerical implementation, we introduce the Majorana operators 
\begin{equation}
   \op{\psi}_{2j+1}=\op{a}^\dagger_j+\op{a}_j\,
   \quad \text{and} \quad
   \op{\psi}_{2j}=i(\op{a}_j-\op{a}^\dagger_j)\,,
\end{equation}
so that $\{\op{\psi}_a,\op{\psi}_b\}=2\delta_{a,b}$. Each Gaussian state is completely fixed by the covariance matrix
\begin{equation}\label{eq:majorana_covariance}
	C_{a b}(\rho)
    =\frac{i}{2} \operatorname{tr}
    \left( \rho\left[\op{\psi}_{a}, \op{\psi}_{b}\right]\right)\,,
\end{equation}
a $2L\times 2L$ real, skew-symmetric matrix. Upon beginning in a Gaussian initial condition $\ket{\Psi_0}$, the state remains Gaussian at all times; its covariance matrix is efficiently time evolved as follows. Consider a Gaussian state $\ket{\Psi_t}$ specified by a covariance matrix $C_t$. Suppose that we want to update the state with a bilinear unitary 
\begin{equation}
	\label{eq:unitary_mat}
	\op{\mathcal{O}}=\exp\left(-\frac{1}{4}\sum_{m,n}K_{m,n}\op{\psi}_m\,\op{\psi}_n\right)\,,
\end{equation}
wherein $K_{m,n}\in\mathbb{R}$ is real and skew-symmetric. By the Baker-Campbell-Haussdorf formula and the Majorana anticommutation relations, $\op{\mathcal{O}}\ket{\Psi_t}$ has the covariance matrix $C_{t+1}=O C_t O^T$ specified by the $2L\times 2L$ matrix $O=\exp(K)\, .$
$K$ is the matrix with elements $K_{m,n}$, as in~\eqref{eq:unitary_mat}. $\op{\mathcal{O}}$ is a $2^L\times2^L$ unitary operator, whereas $O$ is a $2 L \times 2 L$ orthogonal matrix. Therefore, to classically simulate the time evolution, one must only calculate exponentials and products of matrices whose dimensionalities scale linearly with the system size. Hence the whole algorithm is efficient. 

To apply this procedure to our problem, we rewrite the Hamiltonians~\eqref{eq:H1} and~\eqref{eq:H2} in terms of the Majorana operators:
\begin{align}
   \op{H}_1(q)=i \, \frac{\alpha}{2} \,
   \sum_{j=0}^{L/2-1}\op{\psi}_{4j+2q}\op{\psi}_{4j+2q+1} \, ,
   \quad \text{and} \quad
   \op{H}_2=i \, \frac{\pi}{4} \,
   \sum_{j=0}^{L-1}(-1)^{j+1} \op{\psi}_{2j}\op{\psi}_{2j+3} \, .
\end{align}
We have assumed that the boundary conditions are periodic. $\op{H}_2$ may require a specific boundary term depending on the desired initial and boundary conditions~\cite{lieb1961two,cabrera1987role}. Finally, one must specify the initial state's covariance matrix. In our numerical simulations, the initial state is the fermionic vacuum state $\ket{\Omega}$ (which is a  Gaussian state and has definite fermion-number-parity). This state is defined by the condition $\op{a}_j\ket{\Omega}=0$	for all $j=0, 1, \ldots, L-1$. In qubit language, $\op{\sigma}^y_j\ket{\Omega}=\ket{\Omega}$ for all $j$:
\begin{equation}
\ket{\Omega}=\left(\frac{\ket{0}+i\ket{1}}{\sqrt{2}}\right)^{\otimes L}\,.
\end{equation}
Recall the covariance-matrix definition~\eqref{eq:majorana_covariance} and the relationship between fermionic and Majorana operators. Using these inputs, one can compute the covariance matrix immediately: 
$C_{ab}(\ket{\Omega} \bra{\Omega})= \delta_{a-1,b}-\delta_{a+1,b}\,.$

\section{Single-charge generalized Gibbs ensemble}\label{sec:Q1_GGE}

This section reports analytic results concerning the GGE $\op{\rho} = e^{-\mu \op{Q}_1}/Z$ that accounts for the conservation of $\op{Q}_1=\sum^{L-1}_{j=0} \op{\sigma}^z_j \op{\sigma}^z_{j+1}$ alone. We compute the partition function $Z=\Tr(e^{-\mu \op{Q}_1})$, chemical potential $\mu$, and $z$-type expectation values 
$\Tr(\op{\rho} [\op{\sigma}^z]^{\otimes k})$. 
We use the computational basis, defined through
$\op{\sigma}^z_j \ket{m_j} = z_{m_j}\ket{m_j}$, wherein $j=0,\ldots,L-1$, \: $m_j \in \{0,1\}$, \: $z_0=1$, \: and $z_1=-1$.

Define the transfer matrix $P$ with elements $\bra{m_j } P \ket{m_{j+1}}$:
\begin{equation} \label{eq:P}
    P = 
    \left(
\begin{array}{cc}
e^{-\mu}& e^{\mu}\\
e^{\mu}& e^{-\mu}
\end{array}
\right) 
    \,.
\end{equation}
The eigenvalues of $P$ are $\lambda_+=2 \cosh \mu$ and $\lambda_-=-2\sinh \mu$. We introduce $P$ to represent the partition function as $Z = \Tr(e^{-\mu \op{Q}_1}) = \Tr(P^L) = \lambda_+^L+\lambda_-^L$.
The chemical potential is set by the initial condition, here the tilted ferromagnet $\ket{\Psi_0} = \ket{\Psi(\theta, \phi)}$. In the thermodynamic limit, one expects,
\begin{equation}\label{eq:fix_mu}
    \lim_{L \to \infty} \bra{\Psi_0} \op{Q}_1\ket{\Psi_0} = \lim_{L \to \infty} \Tr {\bm{(}}\op{\rho} \op{Q}_1 \bm{)}\, .
\end{equation}
Since $\ket{\Psi_0}$ is translationally invariant, 
$\bra{\Psi_0} \op{Q}_1\ket{\Psi_0}
= L \bra{ \Psi_0 } \op{\sigma}^z_0\op{\sigma}^z_1\ket{\Psi_0}
=L\cos^2 \theta$. 
Similarly, $\Tr(\op{\rho} \op{Q}_1)
=L\Tr(\op{\rho} \op{\sigma}^z_0\op{\sigma}^z_1)$. 
More generally,
\begin{align}\label{eq:zk_exp}
\Tr \left ( \op{\rho} [\op{\sigma}^z]^{\otimes k}\right ) &= \frac{1}{Z} \Tr \left ( [\op{\sigma}^z P]^k P^{L-k}\right ) \\
&= \frac{1+(-1)^k}{2} \left ( \frac{\lambda_+^L \lambda_-^k + \lambda_+^k \lambda_-^L}{\lambda_+^L+\lambda_-^L} \right ) \left [ -2\sinh (2 \mu) \right ]^{-k/2}\,.
\end{align}
If $k=2$, in the thermodynamic limit (as $L \to \infty$), $\lambda_+^L \gg \lambda_-^L$, so $\Tr \left ( \op{\rho} [\op{\sigma}^z]^{\otimes 2}\right ) \to -2 \sinh^2 \mu / \sinh 2 \mu = -\tanh \mu$. Therefore,
\begin{equation} \label{eq:chemical_potential}
    \mu = -{\rm arctanh} \left ( \cos^2 \theta \right ).
\end{equation}
Finally, inserting Eq.~\eqref{eq:chemical_potential} into~\eqref{eq:zk_exp}, we calculate the GGE prediction for the tilted-ferromagnet initial condition:
\begin{equation}\label{eq:gibbs_prediction}
\Tr \left ( \op{\rho} [\op{\sigma}^z]^{\otimes k}\right )= \left( \frac{1+(-1)^k}{2} \right ) \left ( \frac{\cos^k \theta+ \cos^{2L-k} \theta}{1+\cos^{2L} \theta} \right )  \xrightarrow{\mathit{L \to \infty}}  \left ( \frac{1+(-1)^k}{2} \right )\cos^k \theta \, . 
\end{equation}
The GGE predicts that even-$k$ $z$-type expectation values thermalize to their initial values, while odd-$k$ $z$-type expectation values thermalize to zero.

Here we make one comment on GGE predictions made using all the charges~\eqref{eq:charges}. The main text observed in Fig.~\ref{fig:numerics} that, as $k$ increases,
the equilibrium values approaches zero. We understand this trend in the Heisenberg picture as follows. Return to expression~\eqref{eq:correlations} for expectation values $\mathcal{F}(t;,\gamma,k)$.  Consider preparing $\ket{\Psi_0}$, evolving it forward in time to $\ket{\Psi_t}$, perturbing it with $[\op{\sigma}^y]^{\otimes k}$, and evolving backward in time. The resultant state's overlap with the initial state, $\ket{\Psi_0}$, is $\mathcal{F}$. If the perturbation were absent (if $k$ equaled 0), the overlap would be 1. The larger the perturbation, the more the resultant state should differ from the initial state, so the smaller the overlap should be, consistent with our observations.

\section{Gates with larger supports}\label{sec:larger_gates}

An important question is whether our free-fermionic mapping can extend to more-general constrained discrete dynamics. For instance, Ref.~\cite{hillberry2021entangled} reported about QCA gates supported on $>3$ qubits. Interesting behavior was attributed to Goldilocks gates supported on five qubits, four of which formed the control set. More generally, we can consider QCA whose local gates act on $d>2$ neighboring qubits, $2s<d$ of which form the control set.

We could not find integrable QCA that had the form~\eqref{eq:two-step_evolution} and whose gates acted on $>3$ qubits.
%larger support. 
However, exploiting our previous results, one can easily exhibit slightly more-general constrained models that map to free fermions. We will construct examples with arbitrary $d\geq 4$ and $s=1$.  

Consider the $d$-qubit neighborhood gates
\begin{align} \label{eq:Uj_general}
	\op{u}_j & \coloneqq \sum_{m,n=0}^1 \ket{m}\bra{m}_{j}
	\otimes  \op{V}_{j+1,\ldots j+d-2}(m,n)\otimes \ket{n}\bra{n}_{j+d-1}\,.
\end{align}
Qubits $j$ and $j+d-1$ form the control set. However, unlike in Eq.~\eqref{eq:uj}, the update operator $\op{V}(m,n)$ may depend nontrivially on the control qubits' states. Define the partial-step updates
\begin{equation}
\op{G}(q) \coloneqq \prod_{j} \op{u}_{(d-s)j+q} \, ,
\end{equation}
assuming that $(s+d)$ divides $L$. The single-step Floquet operator may be defined as
\begin{equation}
    \label{eq:U_gen}
    \op{U} \coloneqq \prod_{q=0}^{d-2}\op{G}(q)\,.
\end{equation}

We find a parametric class of QCA that satisfy Eqs.~\eqref{eq:Uj_general}--\eqref{eq:U_gen} and that map to free fermions. The local gates have the form 
\begin{equation}
    \op{U}_j=\op{W}^{(1)}_{j+1\ldots ,j+d-3} \prod_{m=j}^{j+d-3}[CZ]_{m,m+1} \, \op{W}^{(2)}_{j+1\ldots ,j+d-3} \, .
\end{equation}
$W^{(1)}$ and $W^{(2)}$ denote arbitrary unitary operators mapped to Gaussian ones via the JW transformation~\eqref{eq:fermionic}. For instance, we can choose
\begin{equation}
\op{W}^{(a)}_{j+1\ldots,j+d-3}=\exp\left(-i\sum_{{\ell}=j+1}^{j+d-3}c_{\ell}^{(a)}\op{\sigma}^{y}_{\ell}
-i\sum_{\gamma,\delta\in\{x,z\}}\sum_{{\ell}=j+1}^{j+d-4}c_{{\ell},\gamma,\delta}^{(a)}\op{\sigma}^{\gamma}_{\ell}\op{\sigma}^{\delta}_{{\ell}+1}\right) \,.
\end{equation}
The $a=1,2$, while $c_{j}^{(a)}$ and $c_{j,\gamma,\delta}^{(a)}$ are arbitrary real parameters. The second sum is over only the Pauli operators $\op{\sigma}^x$ and $\op{\sigma}^{z}$. The nontrivial part of the evolution operator is $\prod_{j}[CZ]_{j,j+1}$. The JW transformation maps this factor to a Gaussian operator~\eqref{eq:fermionic}, by Eq.~\eqref{eq:global_cz}. By construction, the individual operators $\op{W}^{(a)}_{j+1\ldots ,j+k-3}$ is also Gaussian. Therefore, the whole Floquet operator $\op{U}$ [Eq.~\eqref{eq:U_gen}] is mapped to a product of Gaussian operators. Hence $\op{U}$ is Gaussian.

\bibliographystyle{apsrev4-2}.
\bibliography{refs}

\end{document}